%% file: main.tex
\RequirePackage{fix-cm}
\documentclass[smallextended]{svjour3}       %
\smartqed  %
\usepackage{placeins}
\usepackage{multirow}
\usepackage{booktabs}
\usepackage{xcolor}
\usepackage{geometry}
\usepackage{color}
\usepackage{dirtree}
\usepackage{booktabs}
\usepackage{graphicx}
\usepackage{makecell}
\usepackage[table]{xcolor} 
\usepackage[justification=centering]{caption}
\usepackage{xspace}
\usepackage{enumitem}
\usepackage{tikz}
\usepackage{threeparttable}
\usepackage{subcaption}
\usepackage{tabularx}
\usepackage{makecell}
\usepackage{ragged2e}
\usepackage{adjustbox}
\usepackage{enumitem}
\usepackage{listings}
\usepackage{subcaption}
\usepackage{soul}
\usepackage[most]{tcolorbox}
\tcbuselibrary{skins}
\usepackage{pgf}
\usepackage{collcell}
\usepackage[authoryear,round]{natbib}
\usepackage{hyperref}
\usepackage{orcidlink}
\usepackage{amssymb}

\sethlcolor{yellow!35}

\lstdefinelanguage{Kotlin}{
  keywords={
    package,import,class,interface,object,fun,val,var,
    if,else,when,try,catch,finally,throw,
    for,while,do,return,break,continue,
    in,is,as,this,super,
    public,private,protected,internal,
    open,final,override,abstract,
    companion,const,lateinit,data,sealed,enum,
    suspend,inline,reified,tailrec,
    null,true,false
  },
  sensitive=true,
  morecomment=[l]{//},
  morecomment=[s]{/*}{*/},
  morestring=[b]",
}

\definecolor{codeblue}{RGB}{40,70,130}
\definecolor{codegreen}{RGB}{50,120,70}
\definecolor{codered}{RGB}{150,50,50}
\definecolor{codegray}{RGB}{120,120,120}
\definecolor{codepurple}{RGB}{120,40,140}

\lstdefinestyle{icseTinyKotlin}{
  language=Kotlin,
  basicstyle=\ttfamily\footnotesize,
  numbers=left,
  numberstyle=\tiny\color{codegray},
  stepnumber=1,
  numbersep=5pt,
  breaklines=true,
  breakatwhitespace=true,
  columns=fullflexible,
  keepspaces=true,
  showstringspaces=false,
  frame=single,
  xleftmargin=1.2em,
  keywordstyle=\color{codeblue}\bfseries,
  stringstyle=\color{codepurple},
  commentstyle=\color{codegreen}\itshape,
  identifierstyle=\color{black},
  emph={System,out,println},
  emphstyle=\color{codered}
}

\lstdefinestyle{icseTinyPython}{
  language=Python,
  basicstyle=\ttfamily\footnotesize,
  numbers=left,
  numberstyle=\tiny\color{codegray},
  stepnumber=1,
  numbersep=5pt,
  breaklines=true,
  breakatwhitespace=true,
  columns=fullflexible,
  keepspaces=true,
  showstringspaces=false,
  frame=single,
  xleftmargin=1.2em,
  keywordstyle=\color{codeblue}\bfseries,
  stringstyle=\color{codepurple},
  commentstyle=\color{codegreen}\itshape,
  identifierstyle=\color{black},
  emph={self,cls,None,True,False},
  emphstyle=\color{codered}
}

\newtcbox{\rawbadge}[2][]{enhanced, nobeforeafter, 
  colframe=#1!70!black,
  colback=#1!20, coltext=black,
  boxrule=0.6pt, arc=2pt, boxsep=0pt,
  left=3pt, right=3pt, top=2pt, bottom=2pt,
  fontupper=\tiny,
  box align=base,
  #2}

\tcbset{
  rqbox/.style={
    enhanced,
    colback=paleblue!10!white,
    colframe=paleblue!50!black,
    leftrule=4pt,
    rightrule=0pt,
    toprule=0pt,
    bottomrule=0pt,
    arc=0pt,
    outer arc=0pt
  }
}

\newcommand{\projectstructure}{{\rawbadge[badgeblue]{}{STRUCT}}}
\newcommand{\rawsources}{{\rawbadge[badgepink]{}{RAW}}}
\newcommand{\filesummarytwo}{{\rawbadge[green]{}{ROLESUM}}}
\newcommand{\filesummarythree}{{\rawbadge[white]{}{TECSUM}}}

\newcommand{\noranking}{{\rawbadge[gray]{}{NORANK}}}
\newcommand{\bugreports}{{\rawbadge[badgeyellow]{}{BUGSUM}}}

\definecolor{impLow}{RGB}{223,245,223}   %
\definecolor{impMid}{RGB}{175,225,178}   %
\definecolor{impHigh}{RGB}{120,200,135}  %

\NewDocumentCommand{\icon}{o}{%
  \IfNoValueTF{#1}{\phantom{\rule{1.2em}{1.2em}}}{\includegraphics[height=1.2em]{#1}}}

\definecolor{paleblue}{rgb}{0.7,0.8,1.0}  %
\definecolor{palerblue}{rgb}{0.85,0.90,1.0} %
\definecolor{badgeblue}{HTML}{6db1ff}
\definecolor{badgepink}{HTML}{ff80df}
\definecolor{badgepurple}{HTML}{e08fff}
\definecolor{badgered}{HTML}{fe7070}
\definecolor{badgeyellow}{HTML}{fac570}

\newcommand{\rqone}{\textbf{What is the impact of code representations on traditional retrievers for bug localization?~}}
\newcommand{\rqtwo}{\textbf{What is the impact of code representations on LLM-based bug localization?~}}
\newcommand{\rqthree}{\textbf{How does code representation impact post-retrieval ranking?~}}

\usepackage[strict]{changepage}
\usepackage{framed}

\definecolor{formalshade}{rgb}{0.95,0.95,1}
\definecolor{darkblue}{rgb}{0.0, 0.0, 0.55} %

\begin{document}

\title{Retrieval-Oriented Code Representations in Agentic Bug Localization%
}

\author{Genevieve Caumartin\,\orcidlink{0009-0007-2797-6926}\textsuperscript{1,*}      \and
        Tse-Hsun (Peter) Chen\,\orcidlink{0000-0003-4027-0905}\textsuperscript{1}       \and
        Diego Elias Costa\,\orcidlink{0000-0001-7084-2594}\textsuperscript{1}
}

\institute{ Genevieve Caumartin \at \email{genevieve.caumartin@mail.concordia.ca} \and Tse-Hsun (Peter) Chen \at \email{peterc@encs.concordia.ca} \and Diego Elias Costa \at \email{diego.costa@concordia.ca} \and \mbox{} \at \textsuperscript{1}Department of Computer Science and Software Engineering, Gina Cody School of Engineering and Computer Science, Concordia University, Montreal, QC, Canada\\ \textsuperscript{*}Corresponding author }

\date{Received: date / Accepted: date}

\maketitle

\begin{abstract}

LLM-based agents are increasingly being used to support software development, yet their performance in repository-level tasks depends on retrieving the right code context. Existing studies have explored file-level localization using traditional information retrieval over file paths and raw source code. However, the role of textual code representations in retrieval and localization remains underexplored. %

We study file-level bug localization as a representation-driven retrieval problem. Across the Long Code Arena (LCA) and SWE-bench Verified (SWE) datasets, we compare five code representations: file paths, raw source code, and three LLM-generated textual representations. 
Our experiments include lexical, semantic, and LLM-based retrieval, followed by LLM-based post-retrieval ranking. We quantify the cost incurred by a representation through \textit{representation footprint}.
We find that the choice of code representation affects both localization effectiveness and cost. Role-aware summaries outperform file-path representations by up to 40\% Hit@5 while requiring a representation footprint 10.4 to 20.9$\times$ smaller than raw source code. 
Combining complementary representation results and ranking retrieved candidates with an LLM provides further gains of up to 31.9\% and 42.0\%, respectively. 
Overall, role-aware summaries provide the best cost-effectiveness trade-off, while raw source code offers effectiveness in some settings at a significantly higher cost. 
A case study with \textit{Agentless} reveals the utility of our techniques within a well-known pipeline, reaching 94\% Hit@6 on file localization (+4.7\% against the baseline).
Our findings suggest that code representation should be treated as a first-class design choice in agentic localization pipelines, guided by pipeline stage and cost-accuracy requirements.

\end{abstract}

\input{intro}
\input{relatedwork}

\input{methodology}

\input{rq1}

\input{rq2}
\input{rq3}
\input{discussion}

\input{threats}

\input{conclusion}

\section*{Data Availability Statement}
Datasets, all generated databases and results are available publicly~\citep{databases_results}. The code used to run the experiments is available along with comprehensive documentation~\citep{replication}.

\section*{Author Contributions}
All authors contributed to the study conception and design. Material preparation, data collection and analysis were performed by Genevieve Caumartin, and reviewed by the other authors. The first draft of the manuscript was written by Genevieve Caumartin and all authors commented on previous versions of the manuscript. All authors read and approved the final manuscript.

\bibliographystyle{spbasic}      %
\bibliography{references}   %

\end{document}

%% file: intro.tex
\section{Introduction}

Recently, large language models (LLMs) have been increasingly used to support software development tasks~\citep{li2025riseaiteammatessoftware}, 
including bug fixing, feature implementation and code reviews. 
One prominent application area is automated program repair, which typically consists of three phases: fault localization, patch generation and patch validation~\citep{llm4apr}. 
The first phase, fault localization (or bug localization), requires identifying the code locations that are relevant to a bug before a patch can be generated or validated. This phase inherits the challenge of finding the relevant files within a large repository, which often contains hundreds to thousands of files. 
Repository-level localization is demanding because it involves navigating a complex code structure and understanding inter-file dependencies to localize the files where a bug should be fixed~\citep{repo_understanding}. 
File-level retrieval is key: identifying the correct files affected by a bug is one of the first and most critical steps in the debugging process~\citep{Jain_Dora_Sam_Singh_2024}. Accurate file-level localization narrows the search space for subsequent reasoning or patch generation, enabling downstream tools or developers to pinpoint faulty code snippets more efficiently. 
Recent evidence by ~\citet{chang2025bridgingbuglocalizationissue} suggests that improving the bug localization phase can substantially improve program repair effectiveness.

While prior work on bug localization and code search often focuses on improving retrieval algorithms or embedding models~\citep{bug_loc_deep_learning,xie2025swefixertrainingopensourcellms,enhancing_with_dnn}, repository-scale settings introduce a different bottleneck: the form in which code is textually represented to retrievers and downstream LLMs. At this scale, the contents of candidate files can exceed LLM context limits or significantly impact performance: although LLMs have increased their context windows at each generation, their accuracy still falls when we give them too many tokens to reason over~\citep{swe-bench}. 
As a result, retrieval effectiveness is not only a function of the retrieval model, but also of how source code is transformed into retrievable units. 
Understanding the impact of how code is presented to a retriever, such as file paths, raw source code and summaries, is therefore critical for designing effective retrieval pipelines in large repositories.

Given these constraints, we view repository-level bug localization as a context engineering problem. \citet{mohsenimofidi2026contextengineeringaiagents} define context engineering as the \textit{deliberate process of designing, structuring, and providing task-relevant information to LLMs}. 
Related works have explored techniques such as context summarization~\citep{recomp_rag,makharev2025codesummarizationfunctionlevel,bjenet} or hierarchical bug resolution~\citep{xia2024agentlessdemystifyingllmbasedsoftware} to reduce the amount of irrelevant information presented to the model. 
The latter progressively increases the level of detail of the representation used as context, from coarse file paths to detailed code snippets. 
These approaches suggest that effective agentic workflows require not only better models, but also careful decisions about which code context is selected, how it is represented, and when it is exposed. However, existing work has not systematically studied how the textual representation of repository files affects localization effectiveness and cost.

\textbf{Our work.} Given this gap and the importance of localizing the right files, \textbf{we investigate the impact of different code representations at the file-level localization stage}. Inspired by recent research based on hierarchical localization~\citep{xia2024agentlessdemystifyingllmbasedsoftware,chang2025bridgingbuglocalizationissue}, we set out to investigate how different code representations affect retrieval costs and effectiveness at different stages of file localization (i.e. retrieval, ranking). 
We define \textit{code representation} as the textual content that represents the repository files. 
We analyze five different representations: the first two are derived directly from the repository (raw source code, file paths), while the last three consist of generated summaries at different granularity levels. 

Our study is structured in three parts. First, we investigate the impact of applying different code representations to traditional retrievers such as BM25 and dense embeddings. Second, we examine how those representations impact file-level bug localization using LLMs as retrievers. Next, we investigate how different code representations impact LLM-based post-retrieval ranking. To test the applicability of our proposed techniques, we present a case study where we introduce our best performing techniques into the \textit{Agentless} pipeline. We end with a discussion on the complementarity of representations and implications for researchers and practitioners developing agentic bug localization/resolution pipelines.

We evaluate our approaches on two long-context benchmarks: Long Code Arena (LCA) ~\citep{LongCodeArena} and SWE-bench (SWE) ~\citep{swe-bench}, by measuring how accurately, and at what cost, different code representations identify the source files that need to be fixed given a bug description and repository content. 
Our study is designed to answer the following three research questions: 

\begin{itemize}
    \item \textbf{RQ1.} \rqone
    In this RQ, we evaluate our five representations over BM25 and four dense retrievers in terms of efficiency and costs. Although there is no single best representation across retrievers and datasets, we find that role-aware summaries strike the best cost-effectiveness balance.

    \item \textbf{RQ2.} \rqtwo
    This RQ evaluates three smaller LLMs as retrievers. Each model is instructed to retrieve relevant files given an issue description and the repository under four different representations. Role-aware summaries remain the best choice in terms of cost-efficiency with our larger model (32B), while smaller models lack the semantic understanding required to reliably map summaries to relevant files.

    \item \textbf{RQ3.} \rqthree
    In this section, we evaluate ranking with four different representations along with our two best embedding retriever results with $k$=20. Lighter representations show their practicability in a single-prompt ranking setting, retaining efficiency with higher compression.

\end{itemize}

\noindent
Our main contributions are as follows:

\begin{itemize}[leftmargin=*]

    \item \textbf{Impact of code representation.} We evaluate multiple traditional retrievers and LLMs under different textual code representation settings at different levels of abstraction: file paths, raw source code, role-aware natural language summaries, detailed technical natural language summaries and generated bug report summaries.  

    \item \textbf{Complementarity of code representation.} 
    We find that combining the results of different representations increases localization coverage. We use reciprocal rank fusion (RRF) to combine results from representation pairs. We reach Hit@5 of 89.3\% on LCA and 83.4\% on SWE when combining LLM retrieval results.

    \item \textbf{Case study on \textit{Agentless}.} We integrate our post-retrieval ranking with role-aware summaries into the \textit{Agentless} file localization workflow and reach 94\% top-6 localization, a 4.7\% increase over the baseline.

    \item \textbf{Replication package.} We publish our replication package, including all code, prompts and outputs, as well as all generated databases and results to foster reproducibility and further research~\citep{databases_results,replication}.
\end{itemize}

Our findings reveal that no single representation dominates across models, datasets, and pipeline stages. Raw source files provide strong retrieval signals, but their large footprint make them poorly suited to retrieval and ranking by most models. Role-aware summaries provide the best cost-effectiveness tradeoff, compressing source files into a compact representation that preserves file responsibilities and improves alignment with natural-language bug reports. On the other hand, BM25 is lightweight and efficient but depends on the lexical overlap with bug reports. Taken together, our results suggest that code representations should be treated as a first-class design choice in repository-level localization, depending on the pipeline stage and cost-accuracy trade-off.

%% file: relatedwork.tex
\section{Related Work} \label{sec:related_work}

Our work spans three main bodies of literature: information retrieval-based fault localization, textual code representations for repository-scale software engineering tasks, and synthetic queries and document expansion for retrieval. In this section, we review each area and position our contributions.

\subsection{Information Retrieval-Based Fault Localization (IRFL).}
Fault localization is the process of locating elements in a program (e.g., files, methods, lines) that are the cause of a fault~\citep{retrieval_bug_localization}. Prior IRFL work largely assumes a fixed textual representation of code and focuses on improving similarity functions or ranking models.

Previous works such as BugLocator~\citep{buglocator}, use textual similarity between a bug report and the source code using a revised Vector Space Model (rVSM), outperforming previous information retrieval techniques such as Latent Dirichlet Allocation~\citep{lda} and Vector Space Model (VSM)~\citep{retrieval_bug_localization}. ~\citet{improving_bug_localization} introduce BLUiR, which enhances the performance of BugLocator by adding structured information from source code (class and method names) to improve retrieval accuracy. Hybrid approaches such as DNNLOC~\citep{bug_loc_deep_learning} leverage rVSM with Deep Neural Network (DNN) to recommend the potential buggy file for a bug report.   
\citet{pengfei} evaluate the accuracy/efficiency tradeoff of sparse and dense retrievers on three coding tasks: Program Synthesis, Commit Message Generation, and Assertion Generation. They find that HNSW is better suited for very large repositories where the knowledge base exceeds $10^3$ entries, with a minor drop in accuracy. 

In traditional IRFL, the retriever is seen as the primary bottleneck. However, in LLM-based localization pipelines, retrieval is often a first step for space reduction; retrieved candidates are then provided to downstream LLMs with limited context windows where the representation of the code affects what the model can inspect. This shift makes representation a first-class concern that traditional IR evaluations did not need to address. Instead of proposing a different retriever or improving an existing one, we study the impact of how code is presented to a retriever. We find that retrievers are impacted in different ways by representations, and discover complementarities between representations that can be exploited through rank fusion.

\subsection{Textual Code Representations for Repository-scale Tasks.}

Early works on representations of code show that careful representation design might improve retrieval results.~\citet{bluir} show that prioritizing structured code information such as class and method names can boost retrieval accuracy. Repository-level task resolution involves extracting large amounts of data, often overflowing the model's context window. Prompting techniques, such as iterative prompting and prompt chaining~\citep{Rafi_enhancing_fl_code_analysis} allow the model to eventually see the whole context but are costly, given the high number of input tokens.~\citet{recomp_rag} introduce an extractive compressor, which extracts relevant sentences from retrieved documents and an abstractive one, which generates summaries by synthesizing information from multiple sources. 
\citet{jia_retrieve_fit} propose EditSum, a retrieve-and-edit framework for code summarization where a summary from a similar snippet is edited to better match the input code.~\citet{makharev2025codesummarizationfunctionlevel} investigate code summarization beyond function level, exploring how adding class and repository context using LLMs improves summary quality.
\citet{bjenet} use a pre-trained code LLM to generate summaries from source code. They use a cross-encoder to localize bugs using bug reports and summaries. \citet{xia2024agentlessdemystifyingllmbasedsoftware} introduce a hierarchical bug resolution pipeline that moves from file paths, file skeletons (e.g. class and method definitions) to precise edit locations to localizes and solve real-world GitHub issues.

Prior works on code representation largely focuses on a single representation or on summarization quality. Our study investigates how different choices of textual representations of code propagate through the retrieval pipeline, from space reduction to LLM-based ranking. By systematically comparing representations across retrievers, datasets, and pipeline stages, our findings provide insights for future work on representation-aware pipeline design: which representations to generate, when to use them, and how to combine them to balance cost and localization accuracy.

\subsection{Synthetic Queries and Document Expansion for Code Retrieval.}
Query expansion is an idea originating from information retrieval, aiming to reformulate queries to improve matching with relevant documents. Early works focused on pseudo-relevance feedback (PRF)~\citep{ir_amati}, where a set of pseudo-relevant documents where terms that frequently occur but are not present in the original query are extracted to expand queries.~\citet{nogueira2019documentexpansionqueryprediction} with their technique referred to as doc2query, proposes augmenting documents with synthetic queries predicted from the document text to improve retrieval effectiveness. ~\citet{zheng-etal-2020-bert} build a query expansion model by fine-tuning BERT and note significant improvements over the base model. 

Recent work has shown that LLMs can be used to generate synthetic queries or query augmentations that improve matching between natural language inputs and code artifacts.~\citet{query2doc} propose query2doc, which uses LLMs to generate pseudo-documents based on an initial query. They are then used to expand queries for both sparse and dense retrieval methods with no model fine-tuning.~\citet{jagerman2023queryexpansionpromptinglarge} further explore LLM-based query expansion by prompting LLMs with different prompt designs, demonstrating that LLMs yield expansions with improved retrieval performance, when compared with traditional pseudo-relevance feedback techniques.~\citet{rizzo2024assessingqueryexpansion} find that LLMs of different parameter sizes and configurations are effective for generating expanded queries, especially in a zero-shot scenario.

Based on these works, we generate synthetic bug reports from source files as one of our summary-based representations. Instead of expanding the user query at retrieval time, we generate hypothetical bug reports from each source file in a pre-processing stage. Prior work uses synthetic queries to improve matching; we apply this idea to source files by generating hypothetical bug reports as an alternative file representation.

%% file: methodology.tex
\section{Methodology} \label{sec:methodology}

\begin{figure}
    \centering
    \includegraphics[width=0.9\linewidth]{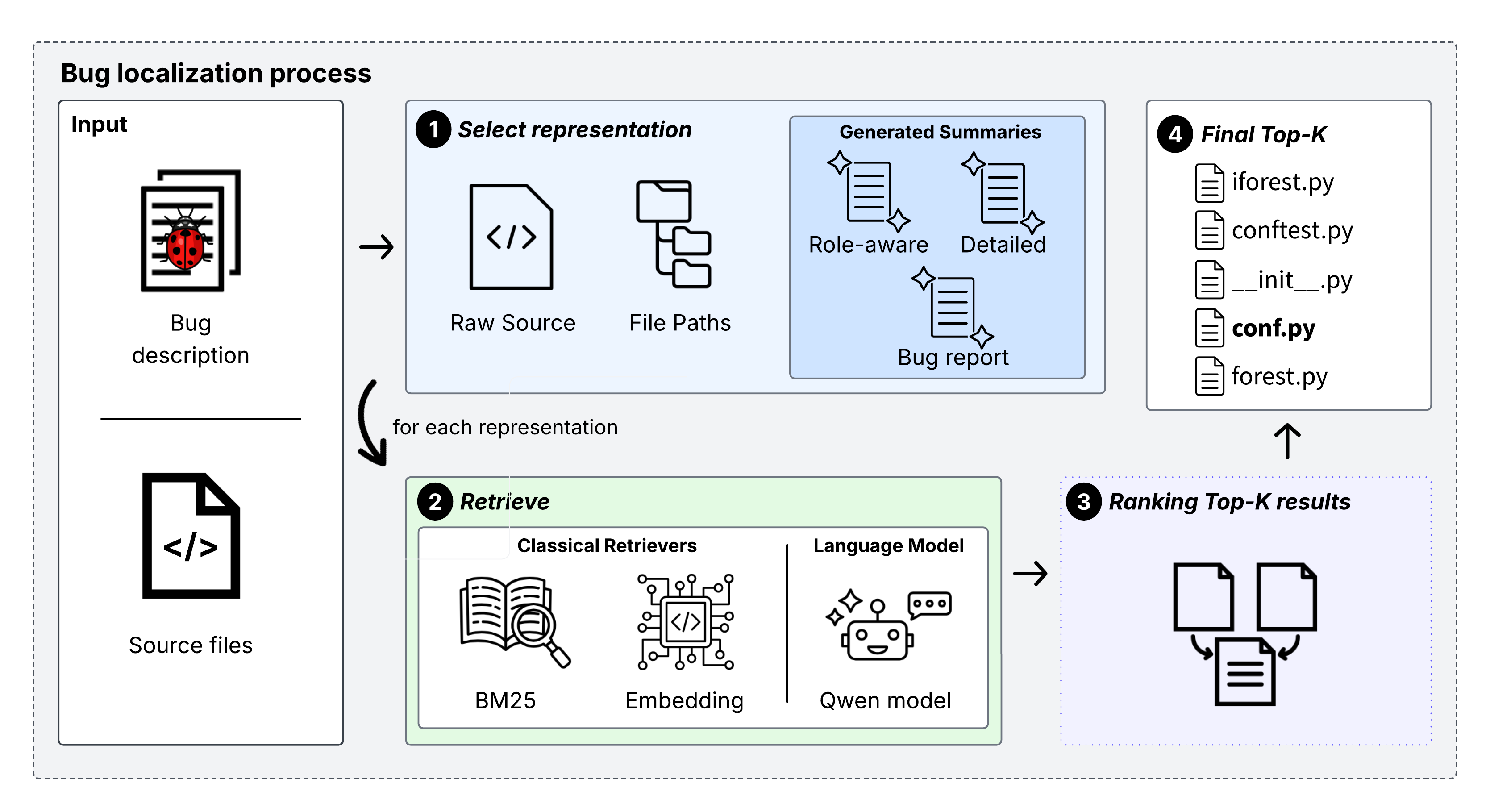}
    \caption{An overview of our file-level retrieval process.} 
    \label{fig:retrieval-process}
\end{figure}

In this section, we describe the building blocks of our study and how they interact to form our file-retrieval process.
Figure~\ref{fig:retrieval-process} illustrates the process, which takes a bug description and the repository source files as input and produces a ranked list of candidate files to fix.
\begin{itemize}
    \item \textbf{Step 1} selects a textual representation of the code, choosing among raw source, file paths, and three types of generated summaries: role-aware, detailed, and bug report. We explain the summary-generation process and these summary types in Section~\ref{sec:code_reprs}.

    \item \textbf{Step 2} retrieves, for each representation, the top-$k$ most relevant files given the bug description, using one of three retrievers: BM25, dense embeddings, or an LLM.

    \item \textbf{Step 3} consolidates the retrieved results into a final ranking, either by fusing the top-$k$ lists of different representations with reciprocal rank fusion~\citep{rrf} or by re-ranking a representation's results with an LLM.

    \item \textbf{Step 4} returns the final top-$k$ files, from which we identify the files relevant to the bug.
\end{itemize}

\subsection{Datasets}
We leverage two datasets that are composed of real-world long-context SE tasks: Long Code Arena~\citep{LongCodeArena} and SWE-bench-Verified~\citep{swe-bench}. SWE-bench is a widely used benchmark in software engineering research~\citep{swe-agent,xia2024agentlessdemystifyingllmbasedsoftware,autocoderover}. We pair it with Long Code Arena, whose Bug Localization subset contains 50\% multi-file tasks. Table~\ref{tab:dataset-stats} summarizes both datasets' statistics.

\input{datasets_stats}

\textbf{Long Code Arena} contains six repository-level benchmarks. We concentrate on the Bug Localization dataset, which features multiple programming languages (Python, Java, Kotlin). We use the test set, which contains a total of 150 manually verified data points for the Python, Java and Kotlin subsets: 50 data points each. In 150 tasks, 50\% require changes in two or more files (besides test files), and the other half are single-file tasks.

\textbf{SWE-bench} tests models' ability to resolve real-world GitHub issues and is exclusively focused on Python. There are three variations of the dataset; we select the SWE-bench Verified (SWE) variation as it is manually verified by experts\footnote{\url{https://openai.com/index/introducing-swe-bench-verified/}} and contains a mix of single and multi-file tasks. In 500 tasks, 85.8\% (429) require changes in a single file, while 14.2\% (71) require changes in 2 or more files.

\subsection{Project Code Representations}\label{sec:code_reprs}
\begin{figure} [h]
    \centering
    \includegraphics[width=1.0\linewidth]{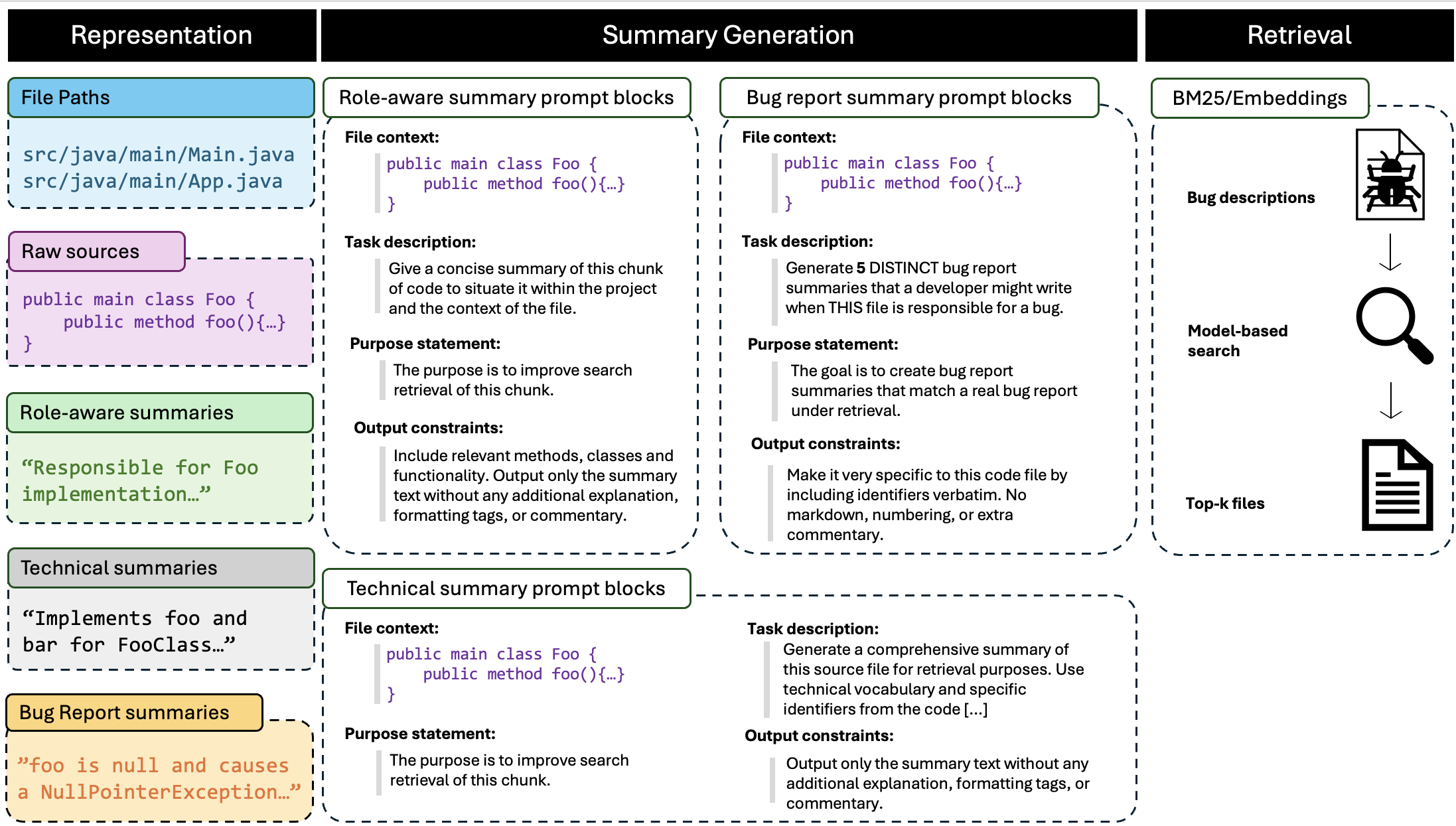}
    \caption{Representations and Prompting Strategies (RQ1).}
    \label{fig:rq1-representations}
\end{figure}

We aim to evaluate different textual code representations to discover how each contributes to retrieval. To understand how they impact lexical and embedding-based retrieval accuracy, we experiment with three different representation types: 

\begin{enumerate}[leftmargin=*]
    \item \textbf{Project structure} (\projectstructure). This representation shows the relative file paths of each file. It is our most lightweight representation, which we use as a baseline.  

    \item \textbf{Raw source files} (\rawsources). Raw source files consist of the repository source files, with no transformation. We use raw source files as a high-fidelity representation that preserves complete lexical and structural information. This choice represents the most resource intensive option, as no compression is applied.

    \item \textbf{File-level generated summaries}. We use LLMs to generate a natural-language summary for each file. We initially experiment with four different prompts with different task granularity levels. We adopt a block-structured prompt template, consistent with empirical analyses of real-world prompt templates that decompose prompts into reusable component types~\citep{prompt_to_templates}. All summary-generation prompts share fixed blocks: (i) a file context block containing the raw source code, (ii) an explicit purpose statement indicating that the summary is intended to support retrieval, and (iii) strict output constraints enforcing summary-only responses.
    We vary the task description block across four prompt variants, ranging from a lightweight structural overview to a detailed technical summary. Sample prompts are shown in Figure~\ref{fig:rq1-representations}. After preliminary experimentation, we select three variations:

    \begin{itemize}[leftmargin=*,noitemsep]
  
  \item \textbf{Role-aware summary (\filesummarytwo).} 
  The summary describes the classes, methods, and functionality, including the file’s role and responsibilities within the codebase. This summary is designed to be concise, while including the file's main classes and methods as well as its role in the repository.
  \item \textbf{Detailed technical summary (\filesummarythree).} 
  The summary includes details covering purpose, APIs, dependencies, error handling, interactions, and design patterns, using precise technical identifiers. This summary is meant to be comprehensive and include code identifiers that are required for lexical retrieval.

    \item \textbf{Bug reports summaries}(\bugreports). Inspired by prior work on document expansion via pseudo-query prediction approximating the vocabulary and intent of real bug reports, we ask an LLM to generate plausible bug reports for each source file~\citep{nogueira2019documentexpansionqueryprediction}. After preliminary experiments, we set the number of bug report summaries generated per file to five. An example of a small file and its generated bug report summaries is shown in Figure~\ref{fig:snippet-queries}.
    \end{itemize}

\end{enumerate}

\textbf{Summary generation.} For summary generation, we select GPT-OSS 20B~\citep{openai2025gptoss120bgptoss20bmodel}, a medium-sized general-purpose model. Given the scale of summary generation across 650 repositories, we prioritize a model accessible via the Groq API\footnote{\url{https://groq.com/}} for its high throughput and cost efficiency. While a code-specific model could be considered, our focus is on the quality of the generated summaries rather than the model used to produce them, and preliminary tests showed satisfactory output quality. 

\input{snippet-queries}

\subsection{Efficiency and Cost Analysis} \label{sec:repr_costs}

Retrieval incurs costs that vary depending on the representation and the retriever used. A very efficient representation-retriever pair that results in poor retrieval results is unlikely to gain adoption, while a heavier but more accurate pair may suit many workflows, as long as the resource requirements remain accessible. Costs can be divided into a few categories:
\begin{itemize}
    \item Context construction: retrieving, ranking, summarizing repository context before repair.
    \item Representation footprint: cost to store each representation.
    \item Indexing and retrieval costs: creating a BM25 index, embedding generation, vector database storage, LLM-based retrieval.
    \item Model inference costs: input/output tokens, API pricing, hardware requirements.
\end{itemize}

We focus on the costs associated with \textbf{context construction}, and \textbf{representation footprint}. Different representations expose different amounts of information to downstream retrievers, but they also differ in pre-processing cost and storage footprint. For example, raw source code provides detailed information but produces a large footprint, while project structure is lightweight but contains limited semantic content. Summary-based representations introduce an additional generation step, which we can measure directly in terms of API cost. We therefore analyze representation footprint and summary-generation cost to characterize the efficiency trade-offs involved in using different representations for file-level localization. Our cost analysis attempts to quantify the cost incurred by each combination of representation and retriever. We aim to guide the selection of an appropriate representation given resources and/or monetary constraints. 
\vspace{0.5em}

\textbf{Context construction.} Summary-based representations (\filesummarytwo, \filesummarythree~and \bugreports) incur an additional one-time extra cost to generate the summaries. This is done by providing a LLM the source files, one by one, and asking it to generate a summary. We call this a \textit{preprocessing cost}. Given this extra cost, we aim to discover whether it is worth it, in terms of efficiency and accuracy, to use summaries for file-level retrieval. We exclude any further updates to the summaries required after source code changes, which are assumed to involve only a subset of files at a time per project. We measure summary generation costs in terms of API costs. We report overall API costs and per repository costs in Section~\ref{sec:summary_costs}.
\vspace{0.5em}

\textbf{Representation footprint.}
We measure \textit{token volume} to estimate representation footprint by computing how many tokens are in the search index for each representation. This footprint applies to all retrievers, and is not meant to estimate overall resource consumption. For RQ1, we compute token volume using a single reference tokenizer for all representations, for comparability. As we are using GPT-OSS-20B~\citep{openai2025gptoss120bgptoss20bmodel} for summary generation, we use that model's tokenizer to compute all token counts. For RQ2, we report \textit{representation footprint} as the number of input tokens passed to each LLM, computed with each model's tokenizer. This count includes both the prompt template and the selected repository representation. Since the prompt template is fixed within each experimental setting, differences in footprint are primarily driven by the representation.

\subsection{Combining retrieved results} 
To improve retrieval coverage, specifically on Hit@5, we investigate an ensemble approach that combines the results obtained with each representation. 
Inspired by research on ensembling methods and fused ranking strategies ~\citep{fox1994combination,rank_fusion_tse,rrf}, we experiment with rank fusion algorithms and further detail our approach in each RQ (Sections~\ref{sec:rq1} and ~\ref{sec:rq2}). To assess whether a combination is beneficial, we obtain percentage improvements by comparing with the best performing representation in each combination.

To combine the ranked results of different representations, we use reciprocal rank fusion (RRF), a state-of-the-art rank fusion approach that ranks documents based on their reciprocal rank~\citep{rrf,mrr}. We report our final results, a single fused top-$k$ list, using RRF.

\subsection{Metrics} \label{sec:metrics}

We use two complementary and widely used metrics from information retrieval: Mean Average Precision (MAP) and Hit@k~\citep{flexfl,LongCodeArena,irfl_best_practices}. 
\begin{itemize}
    \item \textbf{MAP@k} represents the quality of the ranking in retrieval tasks. It measures the average position of all relevant files found by retrieval~\citep{irfl_best_practices,flexfl}. The metric evaluates the ranking quality, accounting for the order of retrieved files. 
    \item \textbf{Hit@k} represents the proportion of rankings that contains at least one valid candidate. The metric informs about the proportion of tasks that include at least one relevant file. 
\end{itemize}

Previous studies report $k$=1, 5 and 10~\citep{improving_bug_localization,buglocator}. To reflect a realistic developer inspection budget in practical scenarios, we report results over the top-5 ranked files ($k$=5).

\subsection{Experimental Environment} \label{sec:env_setup}

For RQ1, we use our desktop computer (Mac Studio M4 Max, 64GB RAM) and laptop (MacBook Pro M3, 16GB RAM) to run BM25 and vector embeddings generation.
The computational requirements of our experiments for RQ2 necessitated the use of multiple computing clusters, including HPC with A100 GPUs, where one to two allocations of 20GB GPU memory were utilized for our experiments.
In addition, all LLM-based experiments are run on the desktop computer at least once.

%% file: datasets_stats.tex
\begin{table}
  \centering
      \caption{Summary statistics for the \textsc{Long Code Arena} and \textsc{SWE-bench} datasets.}
      \label{tab:dataset-stats}
    
      \begin{tabular}{@{}l c cc cc@{}}
        \toprule
        \textbf{Dataset}
          & \textbf{\# Tasks}
          & \multicolumn{2}{c}{\textbf{Total files}} 
          & \multicolumn{2}{c}{\textbf{Buggy files}}\\
          & &\multicolumn{2}{c}{\textbf{per task}}  & \multicolumn{2}{c}{\textbf{per task}} \\
          \cmidrule(lr){3-4} \cmidrule(lr){5-6}
          &
          & {Mean} & {Max}
          & {Mean} & {Max}\\
        \midrule
        Long Code Arena & & &\\
         - Bug Localization
          & 150
          & 559 & 7,917
          & 2 & 21\\
        \midrule
        SWE-bench Verified
          & 500
          & 654 & 1,008
          & 1.7 & 31\\
        \bottomrule
      \end{tabular}

\end{table}

%% file: snippet-queries.tex
\begin{figure}[t]
\caption{Sample Kotlin code snippet and derived bug reports.}
\label{fig:snippet-queries}
\begin{lstlisting}[style=icseTinyKotlin]
class User(val name: String) {
    val defaultFriend = User("Default Bot") 
}

fun main() {
    val me = User("Alice") 
}
\end{lstlisting}

\footnotesize
\setlength{\tabcolsep}{4pt}
\begin{tabularx}{\columnwidth}{@{}p{0.08\columnwidth}X@{}}
\toprule
\textbf{\#} & \textbf{Synthetic bug report} \\
\midrule
Q1 & User.defaultFriend recursively constructs User("Default Bot") without termination. \\

Q2 & main() crashes when executing val me = User("Alice"). \\

Q3 & User("Default Bot") incorrectly receives its own defaultFriend. \\

Q4 & User.defaultFriend creates an unbounded User object chain. \\

Q5 & defaultFriend is instantiated per User instead of being shared. \\
\bottomrule
\end{tabularx}

\end{figure}

%% file: rq1.tex
\section{Repository Representation and Traditional  Retrievers} \label{sec:rq1}

Coding agents localize bugs within repositories that often contain thousands of files, exceeding LLMs' practical context limits.
Traditional retrievers reduce this search space in a cost-effective way~\citep{locagent}, serving as a first step prior to localization.
We evaluate how the code representations outlined in Section~\ref{sec:code_reprs} affect retrieval accuracy under both sparse (BM25) and dense retrieval models, to answer the following question: \textbf{RQ1:} \rqone

\subsection{Traditional Retrievers} 

Our analysis includes two families of retrieval techniques, sparse and dense retrievals. Sparse retrieval refers to a family of classical information‐retrieval techniques that represent both queries and documents as high‐dimensional, term‐based vectors~\citep{Huyen_2024_rag}.
We select one sparse retrieval model in our experimental setup:

\begin{itemize}
    \item \textbf{BM25.} BM25 \citep{bm25} is a sparse document retrieval method widely used for information retrieval tasks \citep{LongCodeArena,swe-bench}. BM25 is short for “Best-Match-25”, a probabilistic term‑weighting function used to rank documents by relevance to a user’s query. 
    We use Pyserini's implementation of BM25 \citep{bm25,Lin_etal_SIGIR2021_Pyserini} with default values for the parameters $k_1=0.9$ and $b=0.4$. 
Pyserini is based on Lucene\footnote{\url{https://lucene.apache.org/}}, a fast indexing engine that can handle millions of files efficiently; thus suitable for large code bases.
A notable advantage of this method is that it runs fast: $\approx5$ minutes to index 150 large repositories, and it does not require specific training, special hardware or usage of a vector database.
\end{itemize}

Dense retrieval uses neural encoders, often based on models like BERT~\citep{bert}, to map both queries and documents into low-dimensional vector embeddings. Instead of matching on exact term overlap, they retrieve documents whose embeddings lie nearest to the query embedding in vector space, capturing the semantics~\citep{Huyen_2024_rag}. 
We select four embedding-based models that are sufficiently lightweight to be deployed on our available hardware. All models show competitive performance on the Massive Text Embedding Benchmark (MTEB)~\footnote{\url{https://huggingface.co/spaces/mteb/leaderboard}}. Specifically, we include:

\begin{itemize}
    \item \textbf{GTE-Large}, a 0.3B-parameter model designed for general text and code embeddings~\citep{gte_large}. This lightweight model can be used with a batch size of 32 on our hardware;
    \item \textbf{Qwen 3}, is a 0.6B-parameter text and code embedding model (Qwen3-Embedding-0.6B)~\citep{qwen3embedding}. We use a batch size of 8 due to the model's memory requirements;
    \item \textbf{CodeXEmbed}~\citep{liu2025codexembed}, a 0.4B-parameter embedding model oriented toward code representations. We use a batch size of 8 due to the model's memory requirements;
    \item \textbf{BGE-Large}~\citep{bge_embedding}, a 0.3B-parameter text embedding model included to assess the performance of a general text embedding model, given that summaries and comments in code are predominantly natural language. This lightweight model can be used with a batch size of 32 on our hardware.
\end{itemize}

All embedding models are accessed through HuggingFace\footnote{\url{https://huggingface.co/}} and used with default hyperparameter settings. Similar to LLMs, they have a limited context window. Following~\citet{xia2024agentlessdemystifyingllmbasedsoftware}, we use token-based chunking to process the raw source files representation. For the summaries representation, given their smaller per-file footprint, we use up to the maximum that fits the context window.

\subsection{Results} \label{sec:rq1_results}

\begin{table*}
\centering
\small
\setlength{\tabcolsep}{2pt}
\caption{MAP@5 and Hit@5 bug localization performance across datasets using different file-level representations in increasing order of token volume. Column '\%'
reports the relative Hit@5 improvement over the \projectstructure~baseline of the same row, in \textcolor{green!60!black}{green} for gains and \textcolor{red!60!black}{red} for losses. Token volume is shown in millions of tokens (MTok) per representation.}
\label{tab:rq1_results}
\resizebox{\textwidth}{!}{%
\input{rq1_retrieval_results_v4}
}
\end{table*}

Table~\ref{tab:rq1_results} shows the performance of each embedding with different representations on the LCA and SWE datasets.
We summarize the main findings below.

\noindent
\textbf{Role-aware (\filesummarytwo) and detailed (\filesummarythree) summaries, along with raw source files (\rawsources) show consistent gains over all models on both datasets.} The highest gain is observed with BM25 on raw source files (LCA: +38\%, SWE: +138\%)). Role-aware summaries perform better than other representations when averaging across all models on LCA. On SWE, raw sources are the most beneficial, although role-aware summaries reach a 55\% improvement with BM25. Surprisingly, with Qwen 3, role-aware summaries only perform equally to the baseline.
\vspace{0.5em}

\noindent
\textbf{Role-aware summaries (\filesummarytwo) are a cost-effective option in most cases.} While it does not always beat raw sources, this representation is effective given its relatively smaller footprint: 10.4$\times$ smaller than raw sources on LCA and 20.9$\times$ smaller on SWE. It remains competitive with raw source files: on LCA, it obtains the best MAP@5 and Hit@5 for GTE-Large and CodeXEmbed.
On SWE, it reaches a better MAP@5 and Hit@5 with GTE-Large  and performs strongly with CodeXEmbed. It improves Hit@5 over the baseline in nine of ten retriever–dataset combinations, with Qwen 3 on SWE being the only tie. 
\vspace{0.5em}

\noindent
\textbf{Detailed summaries (\filesummarythree) are not worth the extra token volume.}
They are approximately 6.2$\times$ larger than role-aware summaries on both datasets. However, they do not consistently outperform the shorter summaries. They are particularly effective with Qwen 3, where they improve over role-aware summaries on both datasets, but GTE-Large, BGE-Large and CodeXEmbed generally perform better with the role-aware, lighter summary.
\vspace{0.5em}

\noindent
\textbf{Bug report summaries (\bugreports) show no benefits across all models and datasets.} This includes both BM25 and embedding-based models, where bug report summaries fail to beat the raw source baseline in any configuration. The gap is particularly extreme with BM25 on SWE, where it reaches only 4.4\% MAP@5 and 9.4\% Hit@5.
\vspace{0.5em}

\noindent
\textbf{File paths (\projectstructure) offer a low-cost alternative.} 
Despite their simplicity and high compression, file paths prove to be more informative than bug report summaries across models and datasets. On SWE-bench, they compare against role-aware summaries in terms of both MAP and Hit@$k$ using Qwen 3 embeddings. 
\vspace{0.51em}

\noindent
\textbf{Token volume is up to two orders of magnitude higher with raw sources (\rawsources) when compared with file paths (\projectstructure).} Although it is a strong performer in most settings, it incurs high costs that need to be paid for each retrieval operation: it is 286.2$\times$ larger than file paths on SWE and 64.3$\times$ larger on LCA.

\vspace{1.0em}

\subsubsection*{Combining Representation Results}

To evaluate how fusion impact the results, we showcase two types of combinations. First, we select the best representations based on their performance: we include all representations except bug report summaries.
We select our best combinations based on two settings: the best performing embeddings textual representations (\projectstructure, \rawsources, \filesummarytwo, \filesummarythree), and lower-cost representations. We identify lower cost representations as ones with smaller representation footprint: \projectstructure, \filesummarytwo~and \filesummarythree.
We show our results in Tables~\ref{tab:rq1_fusion_all} (heavier combinations) and ~\ref{tab:rq1_fusion_light_all} (lighter combinations). 
\vspace{0.5em}

\begin{table}
\centering
\setlength{\tabcolsep}{3pt}
\caption{Performance of Fused Heavier Textual Representations, in increasing order of token volume. For each combination, the percentage reports the relative difference in Hit@5 from the better-performing representation.
Positive and negative differences are shown in \textcolor{green!60!black}{green} and \textcolor{red!60!black}{red}, respectively. Footprints are reported in millions of tokens (MTok).}
\label{tab:rq1_fusion_all}
\input{rq1_fused_ranking_all}
\end{table}

\begin{table}
\centering
\setlength{\tabcolsep}{3pt}
\caption{Performance of Fused Lighter Textual Representations, in increasing order of token volume. For each combination, the percentage reports the relative difference in Hit@5 from the better-performing representation.
Positive and negative differences are shown in \textcolor{green!60!black}{green} and \textcolor{red!60!black}{red}, respectively. Footprints are reported in millions of tokens (MTok).}
\label{tab:rq1_fusion_light_all}
\input{rq1_fused_ranking_light_all}
\end{table}

\noindent
\textbf{Summaries provide complementary signals to raw sources.} Table~\ref{tab:rq1_fusion_all} shows that role-aware summaries provide sizeable increases in Hit@5 with LCA using CodeXEmbed and BGE-Large, while BM25 only obtains a small 1\% gain. On SWE, GTE-Large, Qwen 3 and CodeXEmbed yield interesting gains (4.1 to 8.9\%). BM25 does not benefit from combinations on SWE, but reaches up to 4.1\% with the complementarity of file paths on LCA. Detailed technical summaries reach 10.5 and 15\% improvements on LCA and SWE, respectively. %
\vspace{0.5em}

\noindent
\textbf{Raw sources and file paths combination is the most cost-effective option.} In Table~\ref{tab:rq1_fusion_all}, we observe that file paths only add a slight increase in representation footprint and improve by up to +12.4\% on LCA. On SWE, using the right embedding model matters: GTE-Large and Qwen 3 benefit of 31.9\% and 7.3\% increases, respectively, while it is detrimental to other models.
\vspace{0.5em}

\noindent
\textbf{Lighter representation combinations benefit BM25 across the board,} reaching up to 38\% improvement on LCA and 21.4\% on SWE when combining both summaries, as per Table~\ref{tab:rq1_fusion_light_all}. 
\vspace{0.5em}

\noindent
\textbf{Other models show modest and inconsistent gains.} Table~\ref{tab:rq1_fusion_light_all} shows that on LCA, GTE-Large benefits from the file paths and detailed technical summaries combination, while CodeXEmbed sees a slight increase with both summaries. All other model-combination pairs obtain worse results. On SWE, Qwen 3 benefits from file paths and role-aware summaries (+8.9\%), while the same combination obtains a small 2.9\% increase with GTE-Large. Those two models also see improvements on the file paths and detailed technical summaries combination. Other model-combination pairs perform worse than the best representation in each pair.

\begin{tcolorbox}[rqbox]
\textcolor{paleblue!30!black}
{\textbf{Takeaway.} Raw source code provides a strong baseline across retrievers, but its representation footprint is up to 286.2$\times$ larger than that of file paths. Role-aware summaries offer a cost-effective alternative, remaining competitive in most settings while reducing the footprint by up to 20.9$\times$ relative to raw source. Combining raw source with a complementary representation improves Hit@5 by up to 31.9\% over the stronger representations. In contrast, combinations containing only lightweight representations produce less consistent gains and depend on the retriever and dataset.}

\end{tcolorbox}

%% file: rq1_retrieval_results_v4.tex
\begin{tabular}{l rr rr@{\,}>{\tiny}r rr@{\,}>{\tiny}r rr@{\,}>{\tiny}r rr@{\,}>{\tiny}r}
\toprule
\multirow{2}{*}{\textbf{Retriever}} &
\multicolumn{2}{c}{\textbf{\projectstructure}} &
\multicolumn{3}{c}{\textbf{\filesummarytwo}} &
\multicolumn{3}{c}{\textbf{\bugreports}} &
\multicolumn{3}{c}{\textbf{\filesummarythree}} &
\multicolumn{3}{c}{\textbf{\rawsources}}\\
\cmidrule(lr){2-3} \cmidrule(lr){4-6} \cmidrule(lr){7-9} \cmidrule(lr){10-12} \cmidrule(lr){13-15} &
\multicolumn{1}{c}{MAP$_5$} & \multicolumn{1}{c}{Hit$_5$} &
\multicolumn{1}{c}{MAP$_5$} & \multicolumn{1}{c}{Hit$_5$} & \multicolumn{1}{c}{\%} &
\multicolumn{1}{c}{MAP$_5$} & \multicolumn{1}{c}{Hit$_5$} & \multicolumn{1}{c}{\%} &
\multicolumn{1}{c}{MAP$_5$} & \multicolumn{1}{c}{Hit$_5$} & \multicolumn{1}{c}{\%} &
\multicolumn{1}{c}{MAP$_5$} & \multicolumn{1}{c}{Hit$_5$} & \multicolumn{1}{c}{\%} \\
\midrule
\textbf{LCA Dataset} & & & & & & & & & & & & & &\\
\midrule
BM25       & 0.209 & 0.453 & 0.215 & 0.573 & \textcolor{green!60!black}{+27\%} & 0.082 & 0.233 & \textcolor{red!60!black}{$-$49\%} & 0.260 & 0.613 & \textcolor{green!60!black}{+35\%} & 0.241 & 0.627 & \textcolor{green!60!black}{+38\%} \\
GTE-Large  & 0.212 & 0.493 & 0.264 & 0.633 & \textcolor{green!60!black}{+28\%} & 0.152 & 0.393 & \textcolor{red!60!black}{$-$20\%} & 0.227 & 0.593 & \textcolor{green!60!black}{+20\%} & 0.237 & 0.593 & \textcolor{green!60!black}{+20\%} \\
Qwen~3     & 0.227 & 0.540 & 0.274 & 0.667 & \textcolor{green!60!black}{+24\%} & 0.189 & 0.453 & \textcolor{red!60!black}{$-$16\%} & 0.284 & \textbf{0.680} & \textcolor{green!60!black}{+26\%} & 0.292 & 0.620 & \textcolor{green!60!black}{+15\%} \\
CodeXEmbed & 0.234 & 0.487 & \textbf{0.301} & 0.667 & \textcolor{green!60!black}{+37\%} & 0.182 & 0.413 & \textcolor{red!60!black}{$-$15\%} & 0.289 & 0.640 & \textcolor{green!60!black}{+31\%} & 0.273 & 0.587 & \textcolor{green!60!black}{+21\%} \\
BGE-Large  & 0.229 & 0.500 & 0.258 & 0.607 & \textcolor{green!60!black}{+21\%} & 0.161 & 0.400 & \textcolor{red!60!black}{$-$20\%} & 0.254 & 0.600 & \textcolor{green!60!black}{+20\%} & 0.270 & 0.613 & \textcolor{green!60!black}{+23\%} \\
\midrule
\textbf{Footprint} &
\multicolumn{2}{c}{1.6} & \multicolumn{3}{c}{9.9 {\tiny\textcolor{darkgray}{(6.2$\times$)}}} &
\multicolumn{3}{c}{10.6 {\tiny\textcolor{darkgray}{(6.6$\times$)}}} & \multicolumn{3}{c}{62.2 {\tiny\textcolor{darkgray}{(38.9$\times$)}}} &
\multicolumn{3}{c}{102.9 {\tiny\textcolor{darkgray}{(64.3$\times$)}}} \\
\midrule
\midrule
\textbf{SWE Dataset} & & & & & & & & & & & & & &\\
\midrule
BM25       & 0.191 & 0.284 & 0.280 & 0.440 & \textcolor{green!60!black}{+55\%} & 0.044 & 0.094 & \textcolor{red!60!black}{$-$67\%} & 0.234 & 0.378 & \textcolor{green!60!black}{+33\%} & 0.475 & 0.676 & \textcolor{green!60!black}{+138\%} \\
GTE-Large  & 0.355 & 0.522 & 0.467 & 0.696 & \textcolor{green!60!black}{+33\%} & 0.224 & 0.374 & \textcolor{red!60!black}{$-$28\%} & 0.401 & 0.626 & \textcolor{green!60!black}{+20\%} & 0.397 & 0.526 & \textcolor{green!60!black}{+1\%} \\
Qwen~3     & 0.418 & 0.626 & 0.417 & 0.626 & {0\%} & 0.247 & 0.398 & \textcolor{red!60!black}{$-$36\%} & 0.467 & 0.684 & \textcolor{green!60!black}{+9\%} & \textbf{0.568} & \textbf{0.768} & \textcolor{green!60!black}{+23\%} \\
CodeXEmbed & 0.437 & 0.516 & 0.499 & 0.724 & \textcolor{green!60!black}{+40\%} & 0.235 & 0.374 & \textcolor{red!60!black}{$-$28\%} & 0.467 & 0.670 & \textcolor{green!60!black}{+30\%} & 0.503 & 0.732 & \textcolor{green!60!black}{+42\%} \\
BGE-Large  & 0.339 & 0.504 & 0.451 & 0.664 & \textcolor{green!60!black}{+32\%} & 0.242 & 0.388 & \textcolor{red!60!black}{$-$23\%} & 0.377 & 0.590 & \textcolor{green!60!black}{+17\%} & 0.508 & 0.732 & \textcolor{green!60!black}{+45\%} \\
\midrule
\textbf{Footprint} &
\multicolumn{2}{c}{2.9} & \multicolumn{3}{c}{39.7 {\tiny\textcolor{darkgray}{(13.7$\times$)}}} &
\multicolumn{3}{c}{33.6 {\tiny\textcolor{darkgray}{(11.6$\times$)}}} & \multicolumn{3}{c}{245.0 {\tiny\textcolor{darkgray}{(84.5$\times$)}}} &
\multicolumn{3}{c}{830.0 {\tiny\textcolor{darkgray}{(286.2$\times$)}}} \\
\bottomrule
\end{tabular}

%% file: rq1_fused_ranking_all.tex
\begin{tabular}{@{}l
    rr@{\,}>{\tiny}r
    rr@{\,}>{\tiny}r
    rr@{\,}>{\tiny}r@{}}
\toprule
\multirow{2}{*}{\textbf{Retriever}} &
\multicolumn{3}{c}{\textbf{\rawsources\projectstructure}} &
\multicolumn{3}{c}{\textbf{\rawsources\filesummarytwo}} &
\multicolumn{3}{c}{\textbf{\rawsources\filesummarythree}} \\
\cmidrule(lr){2-4}
\cmidrule(lr){5-7}
\cmidrule(lr){8-10}

& \multicolumn{1}{c}{\textbf{MAP$_5$}}
& \multicolumn{1}{c}{\textbf{Hit$_5$}}
& \multicolumn{1}{c}{\tiny\textbf{\%}}
& \multicolumn{1}{c}{\textbf{MAP$_5$}}
& \multicolumn{1}{c}{\textbf{Hit$_5$}}
& \multicolumn{1}{c}{\tiny\textbf{\%}}
& \multicolumn{1}{c}{\textbf{MAP$_5$}}
& \multicolumn{1}{c}{\textbf{Hit$_5$}}
& \multicolumn{1}{c}{\tiny\textbf{\%}} \\
\midrule

\multicolumn{10}{l}{\textbf{LCA Dataset}} \\
\midrule

BM25
  & 0.266 & 0.653 & \textcolor{green!60!black}{+4.1\%}
  & 0.255 & 0.633 & \textcolor{green!60!black}{+1.0\%}
  & 0.255 & 0.647 & \textcolor{green!60!black}{+3.2\%} \\

GTE-Large
  & 0.244 & 0.620 & \textcolor{green!60!black}{+4.6\%}
  & 0.269 & 0.633 & {0.0\%}
  & 0.245 & 0.613 & \textcolor{green!60!black}{+3.4\%} \\

Qwen~3
  & 0.296 & 0.660 & \textcolor{green!60!black}{+6.5\%}
  & 0.299 & 0.660 & \textcolor{red!60!black}{$-$1.0\%}
  & 0.297 & 0.667 & \textcolor{red!60!black}{$-$1.9\%} \\

CodeXEmbed
  & 0.299 & 0.660 & \textcolor{green!60!black}{+12.4\%}
  & \textbf{0.308} & 0.700 & \textcolor{green!60!black}{+4.9\%}
  & 0.302 & \textbf{0.707} & \textcolor{green!60!black}{+10.5\%} \\

BGE-Large
  & 0.299 & 0.640 & \textcolor{green!60!black}{+4.4\%}
  & 0.299 & 0.653 & \textcolor{green!60!black}{+6.5\%}
  & 0.299 & 0.673 & \textcolor{green!60!black}{+9.8\%} \\

\midrule
\textbf{Footprint}
  & \multicolumn{3}{c}{104.5}
  & \multicolumn{3}{c}{112.8}
  & \multicolumn{3}{c}{165.1} \\

\midrule
\midrule

\multicolumn{10}{l}{\textbf{SWE Dataset}} \\
\midrule

BM25
  & 0.368 & 0.548 & \textcolor{red!60!black}{$-$18.9\%}
  & 0.459 & 0.668 & \textcolor{red!60!black}{$-$1.2\%}
  & 0.453 & 0.640 & \textcolor{red!60!black}{$-$5.3\%} \\

GTE-Large
  & 0.491 & 0.694 & \textcolor{green!60!black}{+31.9\%}
  & 0.504 & 0.758 & \textcolor{green!60!black}{+8.9\%}
  & 0.473 & 0.720 & \textcolor{green!60!black}{+15.0\%} \\

Qwen~3
  & \textbf{0.577} & \textbf{0.824} & \textcolor{green!60!black}{+7.3\%}
  & 0.563 & 0.808 & \textcolor{green!60!black}{+5.2\%}
  & 0.575 & 0.808 & \textcolor{green!60!black}{+5.2\%} \\

CodeXEmbed
  & 0.495 & 0.720 & \textcolor{red!60!black}{$-$1.6\%}
  & 0.440 & 0.762 & \textcolor{green!60!black}{+4.1\%}
  & 0.446 & 0.720 & \textcolor{red!60!black}{$-$1.6\%} \\

BGE-Large
  & 0.459 & 0.664 & \textcolor{red!60!black}{$-$9.3\%}
  & 0.432 & 0.742 & \textcolor{green!60!black}{+1.4\%}
  & 0.425 & 0.688 & \textcolor{red!60!black}{$-$6.0\%} \\

\midrule
\textbf{Footprint}
  & \multicolumn{3}{c}{832.9}
  & \multicolumn{3}{c}{869.7}
  & \multicolumn{3}{c}{1,075.0} \\

\bottomrule
\end{tabular}

%% file: rq1_fused_ranking_light_all.tex
\begin{tabular}{@{}l
    rr@{\,}>{\tiny}r
    rr@{\,}>{\tiny}r
    rr@{\,}>{\tiny}r@{}}
\toprule
\multirow{2}{*}{\textbf{Retriever}} &
\multicolumn{3}{c}{\textbf{\projectstructure\filesummarytwo}} &
\multicolumn{3}{c}{\textbf{\projectstructure\filesummarythree}} &
\multicolumn{3}{c}{\textbf{\filesummarytwo\filesummarythree}} \\
\cmidrule(lr){2-4}
\cmidrule(lr){5-7}
\cmidrule(lr){8-10}

& \multicolumn{1}{c}{\textbf{MAP$_5$}}
& \multicolumn{1}{c}{\textbf{Hit$_5$}}
& \multicolumn{1}{c}{\tiny\textbf{\%}}
& \multicolumn{1}{c}{\textbf{MAP$_5$}}
& \multicolumn{1}{c}{\textbf{Hit$_5$}}
& \multicolumn{1}{c}{\tiny\textbf{\%}}
& \multicolumn{1}{c}{\textbf{MAP$_5$}}
& \multicolumn{1}{c}{\textbf{Hit$_5$}}
& \multicolumn{1}{c}{\tiny\textbf{\%}} \\
\midrule

\multicolumn{10}{l}{\textbf{LCA Dataset}} \\
\midrule

BM25
  & 0.215 & 0.573 & \textcolor{green!60!black}{$+$30.2\%}
  & 0.261 & 0.567 & \textcolor{green!60!black}{$+$28.9\%}
  & 0.241 & 0.607 & \textcolor{green!60!black}{$+$38.0\%} \\

GTE-Large
  & 0.242 & 0.600 & \textcolor{red!60!black}{$-$5.2\%}
  & 0.220 & 0.607 & \textcolor{green!60!black}{$+$2.4\%}
  & 0.242 & 0.600 & \textcolor{red!60!black}{$-$5.2\%} \\

Qwen~3
  & 0.262 & 0.620 & \textcolor{red!60!black}{$-$7.0\%}
  & 0.248 & 0.620 & \textcolor{red!60!black}{$-$8.8\%}
  & 0.272 & 0.660 & \textcolor{red!60!black}{$-$2.9\%} \\

CodeXEmbed
  & 0.261 & 0.567 & \textcolor{red!60!black}{$-$15.0\%}
  & 0.244 & 0.567 & \textcolor{red!60!black}{$-$11.4\%}
  & \textbf{0.279} & \textbf{0.673} & \textcolor{green!60!black}{$+$0.9\%} \\

BGE-Large
  & 0.238 & 0.593 & \textcolor{red!60!black}{$-$2.3\%}
  & 0.225 & 0.560 & \textcolor{red!60!black}{$-$7.7\%}
  & 0.247 & 0.593 & \textcolor{red!60!black}{$-$2.3\%} \\

\midrule
\textbf{Footprint}
  & \multicolumn{3}{c}{11.5}
  & \multicolumn{3}{c}{63.8}
  & \multicolumn{3}{c}{72.1} \\

\midrule
\midrule

\multicolumn{10}{l}{\textbf{SWE Dataset}} \\
\midrule

BM25
  & 0.286 & 0.450 & \textcolor{green!60!black}{$+$2.3\%}
  & 0.309 & 0.478 & \textcolor{green!60!black}{$+$8.6\%}
  & 0.348 & 0.534 & \textcolor{green!60!black}{$+$21.4\%} \\

GTE-Large
  & 0.429 & \textbf{0.716} & \textcolor{green!60!black}{$+$2.9\%}
  & 0.408 & 0.650 & \textcolor{green!60!black}{$+$3.8\%}
  & 0.387 & 0.680 & \textcolor{red!60!black}{$-$2.3\%} \\

Qwen~3
  & 0.430 & 0.682 & \textcolor{green!60!black}{$+$8.9\%}
  & \textbf{0.440} & 0.692 & \textcolor{green!60!black}{$+$1.2\%}
  & 0.395 & 0.662 & \textcolor{red!60!black}{$-$3.2\%} \\

CodeXEmbed
  & 0.403 & 0.708 & \textcolor{red!60!black}{$-$2.2\%}
  & 0.396 & 0.656 & \textcolor{red!60!black}{$-$2.1\%}
  & 0.390 & 0.686 & \textcolor{red!60!black}{$-$5.2\%} \\

BGE-Large
  & 0.388 & 0.644 & \textcolor{red!60!black}{$-$3.0\%}
  & 0.365 & 0.580 & \textcolor{red!60!black}{$-$12.7\%}
  & 0.374 & 0.652 & \textcolor{red!60!black}{$-$1.8\%} \\

\midrule
\textbf{Footprint}
  & \multicolumn{3}{c}{42.6}
  & \multicolumn{3}{c}{247.9}
  & \multicolumn{3}{c}{284.7} \\

\bottomrule
\end{tabular}

%% file: rq2.tex
\section{Repository Representation and Large Language Models} \label{sec:rq2}

In this section, we study how code representations impact the performance and costs of LLMs as retrievers. 
Beyond traditional lexical and embedding-based retrieval, LLMs introduce a different localization mechanism: rather than matching tokens or embedding vectors, they reason over code semantics, integrate general programming knowledge from pretraining, and can act as retrievers and rankers~\citep{flexfl}. 
Consequently, the impact of code representation may differ from the results we obtained on traditional retrievers (RQ1), as LLM-powered bug localization requires constructing an input context that balances relevance, completeness, and token budget.
Due to resource constraints, we choose to investigate the more lightweight role summaries and leave out the detailed technical summaries (\filesummarythree) for a future study. We study all other representations described in Section~\ref{sec:methodology}, to answer: \rqtwo

\subsection{Retrieval Prompts} 
\label{sec:context_construction}
\begin{figure}
    \centering
    \includegraphics[width=0.75\linewidth]{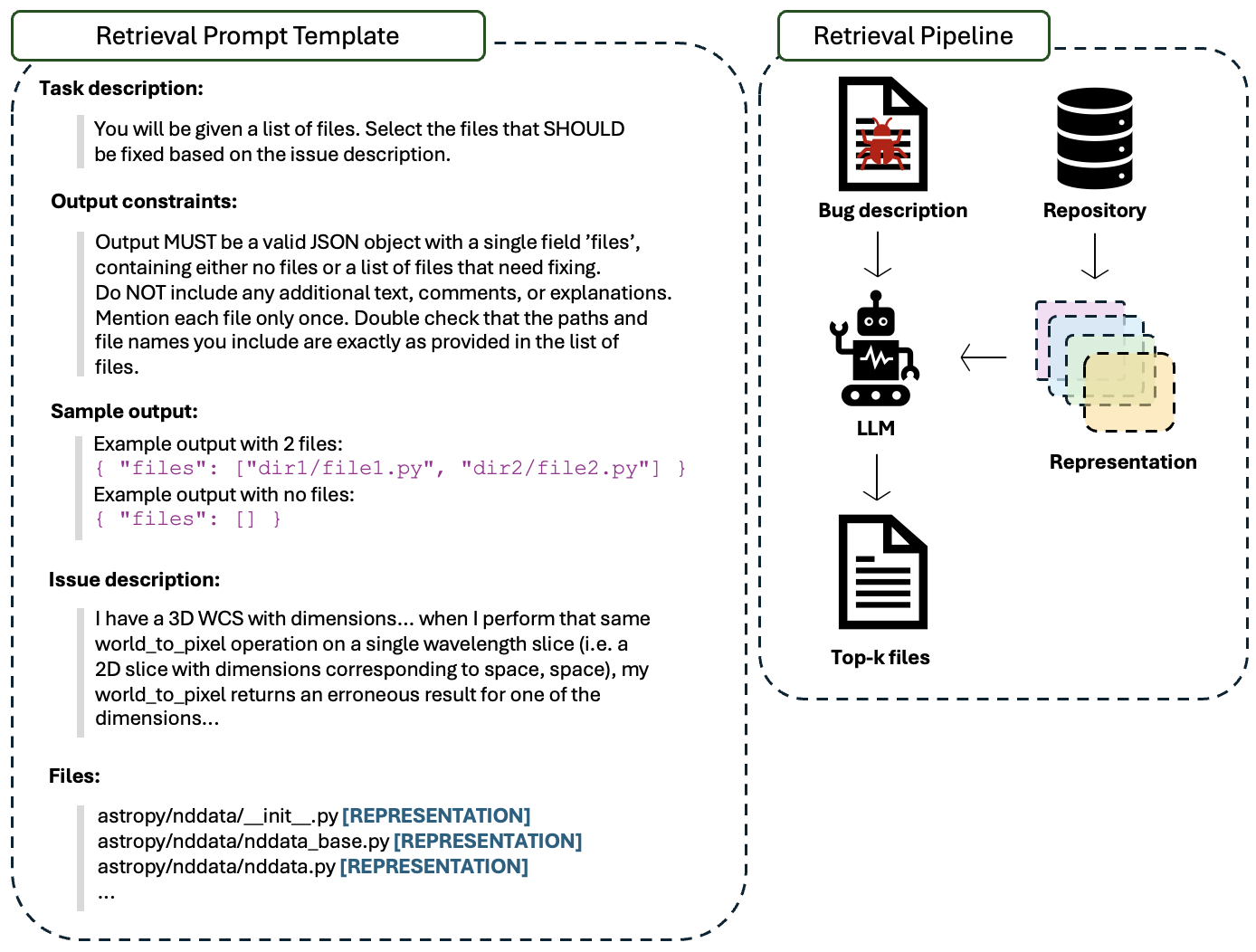}
    \caption{Sample Retrieval Prompt Template. ``[REPRESENTATION]'' is a placeholder for any representation under study.}
    \label{fig:retrieval_prompt_template}
\end{figure}

We design a single prompt template that we adapt to each representation. 
We divide our prompts into distinct blocks, as shown in Figure~\ref{fig:retrieval_prompt_template}.
The issue description and file list are enclosed between tags to avoid any confusion for the model.

We retrieve the issue description from each dataset record, and fetch the associated relative file paths in alphabetical order. Depending on the representation being processed, we may retrieve a summary from our database, given the database key (a unique commit identifier).
We traverse the repository, appending each file. Once the context window is filled, we continue prompting with the remaining files and merge the results in order. In the case where a single file does not fit into the context (e.g. raw sources representation), we truncate it.

\subsection{Large Language Models}

We select models from the Qwen2.5-coder family~\citep{qwen2.5-coder}. 
These models have been specifically trained for code-related tasks, and we evaluate models of three different sizes: 7B, 14B and 32B. All models are prompted using greedy decoding (temperature=0) for maximum output stability. After preliminary testing with 8k, 16k and 32k context sizes, we choose 16k as our fixed context length for all experiments.

In this study, we evaluate smaller open source model and investigate their cost-effectiveness tradeoffs over different representations of source code. Open source models can be used on a local system, do not have privacy issues and mitigate the high costs of using an API~\citep{flexfl}.

\subsection{Results}

\begin{table*}
  \centering
  \setlength{\tabcolsep}{2pt}
  \footnotesize
  \caption{Impact of repository representation on LLM retrieval shown in increasing order of token volume. Percentages report the relative Hit$_5$ difference from the baseline \projectstructure~for the same model and dataset. Positive and negative differences are shown in \textcolor{green!60!black}{green} and \textcolor{red!60!black}{red}, respectively.}
  \label{tab:llm-summary}
  \resizebox{\textwidth}{!}{%
  \input{rq2_results_new}
  }
\end{table*}

We share our results for each representation, per model, in Table \ref{tab:llm-summary}, and summarize the most important findings below. 
\vspace{0.5em}

\noindent
\textbf{The best representation depends on both model size and dataset.}
Qwen 7B achieves its best Hit@5 improvement with raw source on LCA (+53.1\%), but raw source decreases Hit@5 by 8.2\% on SWE, where project structure remains the strongest representation. Qwen 14B exhibits a more consistent preference for raw source, improving Hit@5 by 5.5\% on LCA and 31.3\% on SWE. In contrast, role-aware summaries underperform file paths for Qwen 14B on both datasets, indicating that their usefulness does not increase consistently with model size.
\vspace{0.5em}

\noindent
\textbf{Role-aware summaries (\filesummarytwo) provide the best cost-effectiveness trade-off for Qwen 32B.}
Their benefits are inconsistent for smaller models: Qwen 7B improves by 21.7\% on LCA but decreases by 18.9\% on SWE, while Qwen 14B experiences reductions on both datasets. Qwen 32B, however, consistently improves its Hit@5 with role-aware summaries, reaching gains of 11.6\% on LCA and 10.2\% on SWE. These gains are achieved with approximately 5.4 to 12.2$\times$ the baseline input tokens, compared with 47.6 to 259.5$\times$ times for raw source. %
\vspace{0.5em}

\noindent
\textbf{Raw source (\rawsources) often improves retrieval effectiveness, but at a substantial input token cost.}
Raw sources yield the strongest results for Qwen 14B and the largest LCA improvement for Qwen 7B. However, it requires up to 259.5$\times$ the baseline input tokens and does not consistently outperform lighter representations. In particular, role-aware summaries outperform raw source in Hit@5 for Qwen 32B on both datasets while requiring substantially fewer input tokens.

\vspace{0.5em}

\noindent
\textbf{Bug-report summaries (\bugreports) consistently underperform file paths (\projectstructure).}
They reduce Hit@5 for every model and dataset, with particularly large losses for Qwen 7B and Qwen 14B. Qwen 32B obtains higher absolute scores than the smaller models, but bug report summaries still reduce Hit@5 by 26.2\% on LCA and 31.7\% on SWE. This is consistent with the RQ1 results and suggests that synthetic bug reports are poorly aligned with file retrieval in their current form.

\vspace{1em}

\subsubsection*{Combining representations} 
We select combinations from the three best performing representations with Qwen2.5-coder 32B:~\projectstructure,~\filesummarytwo~and~\rawsources. Combination results are compared against the best contender of each pair. We show our results in Table~\ref{tab:rq2_fusion_results}.
\vspace{0.5em}

\begin{table*}
\centering
\setlength{\tabcolsep}{2pt}
\footnotesize
\caption{Performance of Fused Qwen 32B Textual Representations, in increasing order of token volume. For each combination, the
percentage reports the relative difference in Hit@5 from the
better-performing representation. Positive and negative differences
are shown in \textcolor{green!60!black}{green} and \textcolor{red!60!black}{red}, respectively.}
\label{tab:rq2_fusion_results}
  \resizebox{\textwidth}{!}{%
  \input{rq2_fused_ranking}
  }
\end{table*}

\noindent
\textbf{Most combinations yield improvements over their best single representation.} Heavier representations including raw sources see improvements of up to 38.3\% on Hit@5. Raw sources and file paths are the only exception with Qwen 7B where they tie with the single raw source representation on Hit@5, although MAP@5 improves (+10.0\%). Lighter representations see a similar trend, improving up to 31\% on LCA with Qwen 7B. Qwen 14B is an exception with a decrease of 16.7\%, which aligns with our observation from Table~\ref{tab:llm-summary} where this model performs poorly with summaries.
\vspace{0.5em}

\noindent
\textbf{Raw sources (\rawsources) and file paths (\projectstructure) form the most consistently effective combination.}
This combination improves or maintains Hit@5 in every model and dataset setting and obtains the largest gain in four of the six settings. It is particularly effective for Qwen Coder 14B, improving Hit@5 by 23.2\% on LCA and 11.3\% on SWE. Although Qwen Coder 14B remains below the 32B variant using the same combination on LCA, its combined results surpass all Qwen Coder 32B individual representation results on both datasets. This indicates that complementary representations can partly compensate for reduced model scale.
\vspace{0.5em}

\noindent
\textbf{Role-aware summaries (\filesummarytwo) and file paths (\projectstructure) offer the best cost-effectiveness trade-off.}
This combination requires only approximately 94.5k input tokens on LCA and 314-320k on SWE, compared with approximately 719-877k and 6000-6500 million tokens, respectively, for combinations containing raw source. Despite this substantially smaller context, it improves Hit@5 in five of six settings and achieves the highest MAP@5 for Qwen Coder 32B on SWE. For the same model, it also improves Hit@5 by 11.2\% on LCA, although raw source combinations retain higher Hit@5 on SWE.
\vspace{0.5em}

\begin{tcolorbox}[rqbox]
\textcolor{paleblue!30!black}
{\textbf{Takeaway.} There is no single best representation for all settings: performance depends on both model size and dataset. Role-aware summaries provide the best cost effectiveness trade-off for the largest model, yielding substantial retrieval gains at a fraction of the input token cost of raw source code. Raw sources remains a strong representation, but incurs considerably higher input token costs. Combining representation results is beneficial in most settings and can substantially improve smaller models, although Qwen Coder 32B generally achieves the best performance.}
\end{tcolorbox}

%% file: rq2_results_new.tex
\begin{threeparttable}
\begin{tabular}{@{}l
    rrr
    rr@{\,}>{\tiny}r r
    rr@{\,}>{\tiny}r r
    rr@{\,}>{\tiny}r r@{}}
\toprule
\multirow{2}{*}{\textbf{Model}}
& \multicolumn{3}{c}{\textbf{\projectstructure}}
& \multicolumn{4}{c}{\textbf{\filesummarytwo}}
& \multicolumn{4}{c}{\textbf{\bugreports}}
& \multicolumn{4}{c}{\textbf{\rawsources}} \\
\cmidrule(lr){2-4}
\cmidrule(lr){5-8}
\cmidrule(lr){9-12}
\cmidrule(lr){13-16}

& \multicolumn{1}{c}{\textbf{MAP$_5$}}
& \multicolumn{1}{c}{\textbf{Hit$_5$}}
& \multicolumn{1}{c}{\textbf{FP}}

& \multicolumn{1}{c}{\textbf{MAP$_5$}}
& \multicolumn{1}{c}{\textbf{Hit$_5$}}
& \multicolumn{1}{c}{\tiny\textbf{\%}}
& \multicolumn{1}{c}{\textbf{FP}}

& \multicolumn{1}{c}{\textbf{MAP$_5$}}
& \multicolumn{1}{c}{\textbf{Hit$_5$}}
& \multicolumn{1}{c}{\tiny\textbf{\%}}
& \multicolumn{1}{c}{\textbf{FP}}

& \multicolumn{1}{c}{\textbf{MAP$_5$}}
& \multicolumn{1}{c}{\textbf{Hit$_5$}}
& \multicolumn{1}{c}{\tiny\textbf{\%}}
& \multicolumn{1}{c}{\textbf{FP}} \\
\midrule

\multicolumn{16}{l}{\textbf{LCA}} \\
\midrule

Qwen 7B
& 0.194 & 0.318 & 13.7 (1$\times$)
& 0.185 & 0.387
  & \textcolor{green!60!black}{+21.7\%} & 5.9$\times$
& 0.147 & 0.267
  & \textcolor{red!60!black}{$-$16.0\%} & 6.3$\times$
& \textbf{0.211} & \textbf{0.487}
  & \textcolor{green!60!black}{+53.1\%} & 55.2$\times$ \\

Qwen 14B
& 0.300 & 0.600 & 13.7 (1$\times$)
& 0.242 & 0.480
  & \textcolor{red!60!black}{$-$20.0\%} & 5.9$\times$
& 0.197 & 0.353
  & \textcolor{red!60!black}{$-$41.2\%} & 6.3$\times$
& \textbf{0.305} & \textbf{0.633}
  & \textcolor{green!60!black}{+5.5\%} & 58.1$\times$ \\

Qwen 32B
& 0.344 & 0.687 & 14.8 (1$\times$)
& \textbf{0.353} & \textbf{0.767}
  & \textcolor{green!60!black}{+11.6\%} & 5.4$\times$
& 0.248 & 0.507
  & \textcolor{red!60!black}{$-$26.2\%} & 5.8$\times$
& 0.298 & 0.753
  & \textcolor{green!60!black}{+9.6\%} & 47.6$\times$ \\

\midrule
\midrule

\multicolumn{16}{l}{\textbf{SWE}} \\
\midrule

Qwen 7B
& {0.301} & {0.392} & 26.4 (1$\times$)
& 0.271 & 0.318
  & \textcolor{red!60!black}{$-$18.9\%} & 11.0$\times$
& 0.077 & 0.096
  & \textcolor{red!60!black}{$-$75.5\%} & 10.0$\times$
& 0.280 & 0.360
  & \textcolor{red!60!black}{$-$8.2\%} & 227.5$\times$ \\

Qwen 14B
& 0.444 & 0.512 & 28.3 (1$\times$)
& 0.374 & 0.440
  & \textcolor{red!60!black}{$-$14.1\%} & 10.3$\times$
& 0.174 & 0.216
  & \textcolor{red!60!black}{$-$57.8\%} & 9.4$\times$
& {0.484} & {0.672}
  & \textcolor{green!60!black}{+31.3\%} & 216.9$\times$ \\

Qwen 32B
& \textbf{0.582} & 0.668 & 23.8 (1$\times$)
& 0.501 & \textbf{0.736}
  & \textcolor{green!60!black}{+10.2\%} & 12.2$\times$
& 0.326 & 0.456
  & \textcolor{red!60!black}{$-$31.7\%} & 11.1$\times$
& 0.460 & 0.721
  & \textcolor{green!60!black}{+7.9\%} & 259.5$\times$ \\

\bottomrule
\end{tabular}

\begin{tablenotes}[flushleft]
\footnotesize
\item FP (Footprint): For the baseline, we report average
input tokens in thousands. For other representations, we report
the multiple relative to baseline for the same model and dataset.
\end{tablenotes}
\end{threeparttable}

%% file: rq2_fused_ranking.tex
\footnotesize
\setlength{\tabcolsep}{2pt}
\renewcommand{\arraystretch}{1.05}

\begin{tabular}{@{}l
    rr@{\,}>{\tiny}r r
    rr@{\,}>{\tiny}r r
    rr@{\,}>{\tiny}r r@{}}
\toprule
\multirow{2}{*}{\textbf{Model}}
& \multicolumn{4}{c}{\textbf{\filesummarytwo\projectstructure}}
& \multicolumn{4}{c}{\textbf{\rawsources\projectstructure}}
& \multicolumn{4}{c}{\textbf{\rawsources\filesummarytwo}} \\
\cmidrule(lr){2-5}
\cmidrule(lr){6-9}
\cmidrule(lr){10-13}

& \multicolumn{1}{c}{\textbf{MAP$_5$}}
& \multicolumn{1}{c}{\textbf{Hit$_5$}}
& \multicolumn{1}{c}{\tiny\textbf{\%}}
& \multicolumn{1}{c}{\textbf{Footprint}}

& \multicolumn{1}{c}{\textbf{MAP$_5$}}
& \multicolumn{1}{c}{\textbf{Hit$_5$}}
& \multicolumn{1}{c}{\tiny\textbf{\%}}
& \multicolumn{1}{c}{\textbf{Footprint}}

& \multicolumn{1}{c}{\textbf{MAP$_5$}}
& \multicolumn{1}{c}{\textbf{Hit$_5$}}
& \multicolumn{1}{c}{\tiny\textbf{\%}}
& \multicolumn{1}{c}{\textbf{Footprint}} \\
\midrule

\multicolumn{13}{l}{\textbf{LCA Dataset}} \\
\midrule

Qwen Coder 7B
& 0.247 & 0.507 & \textcolor{green!60!black}{+31.0\%} & 94.5
& 0.232 & 0.487 & 0.0\% & 769.9
& 0.228 & 0.500 & \textcolor{green!60!black}{+2.7\%} & 837.1 \\

Qwen Coder 14B
& 0.198 & 0.500 & \textcolor{red!60!black}{$-$16.7\%} & 94.5
& 0.386 & 0.780 & \textcolor{green!60!black}{$+$23.2\%} & 809.7
& 0.362 & 0.680 & \textcolor{green!60!black}{$+$7.4\%} & 876.8 \\

Qwen Coder 32B
& 0.416 & 0.853 & \textcolor{green!60!black}{+11.2\%} & 94.7
& \textbf{0.427} & \textbf{0.893} & \textcolor{green!60!black}{+18.6\%} & 719.3
& 0.424 & 0.873 & \textcolor{green!60!black}{+13.8\%} & 784.4 \\

\midrule
\midrule

\multicolumn{13}{l}{\textbf{SWE Dataset}} \\
\midrule

Qwen Coder 7B
& 0.405 & 0.490 & \textcolor{green!60!black}{$+$25.0\%} & 316.8
& 0.425 & 0.542 & \textcolor{green!60!black}{$+$38.3\%} & 6032.4
& 0.392 & 0.472 & \textcolor{green!60!black}{$+$31.1\%} & 6296.4 \\

Qwen Coder 14B
& 0.518 & 0.682 & \textcolor{green!60!black}{$+$1.5\%} & 319.8
& 0.600 & 0.748 & \textcolor{green!60!black}{$+$11.3\%} & 6166.6
& 0.555 & 0.726 & \textcolor{green!60!black}{$+$8.0\%} & 6429.8 \\

Qwen Coder 32B
& \textbf{0.653} & 0.802 & \textcolor{green!60!black}{+9.0\%} & 314.2
& 0.552 & 0.816 & \textcolor{green!60!black}{+13.2\%} & 6199.9
& 0.648 & \textbf{0.834} & \textcolor{green!60!black}{+13.3\%} & 6466.5 \\

\bottomrule
\end{tabular}

%% file: rq3.tex
\section{Post-Retrieval Ranking and Representations}\label{sec:rq3}
In this section, we examine the impact of code representation on post-retrieval ranking. 
Ranking allows reordering the selected files according to their importance in fixing a bug, and given that it is often performed on a small subset of files, we hypothesize that code representations may be beneficial without incurring the high costs of the retrieval stage. 
We aim to answer the following research question: \textbf{RQ3.} \rqthree

\subsection{Ranking Settings}
We select our best performing model from RQ2, Qwen2.5-coder 32B, as the ranker for this research question. This model has proven to be the most capable with all representations, yielding the strongest results with both datasets under study. In terms of retrieved results, we select two of our best embedding models from RQ1: Qwen~3 embeddings and CodeXEmbed. 

\textbf{Representation Selection.}
We evaluate four ranking representations that expose distinct forms of information: \projectstructure, \filesummarytwo, \rawsources, and \bugreports. Although the latter was not retained in RQ2 because of its weak standalone retrieval performance, post-retrieval ranking serves a different purpose: it reorders an already retrieved candidate set and may benefit from query-aligned signals that are insufficient for first-stage retrieval. We omit detailed file summaries because they substantially increase representation footprint with less benefits when compared with role-aware summaries.

\textbf{Candidate Selection and baseline.}
We choose to over-retrieve to rank, so matches located towards the end of the list can be brought forward by ranking. We find that Qwen 3 embeddings with~\filesummarytwo~reach 83\% localization on Hit@20. Thus, we select $k$=20 as the number of matches over which we will perform the ranking step. This gives us a generous amount of results, in which most of the ground truth matches are retrieved. For each retriever and each task, we retain up to the top 20 files retrieved in previous RQs using the \projectstructure~representation. This candidate set is fixed across all ranking representations, ensuring that differences in effectiveness result from reordering the same files rather than retrieving different candidates. From this, we create our baseline which we refer to as the \noranking~baseline.

\subsection{Ranking Prompts}

\begin{figure}
    \centering
    \includegraphics[width=0.75\linewidth]{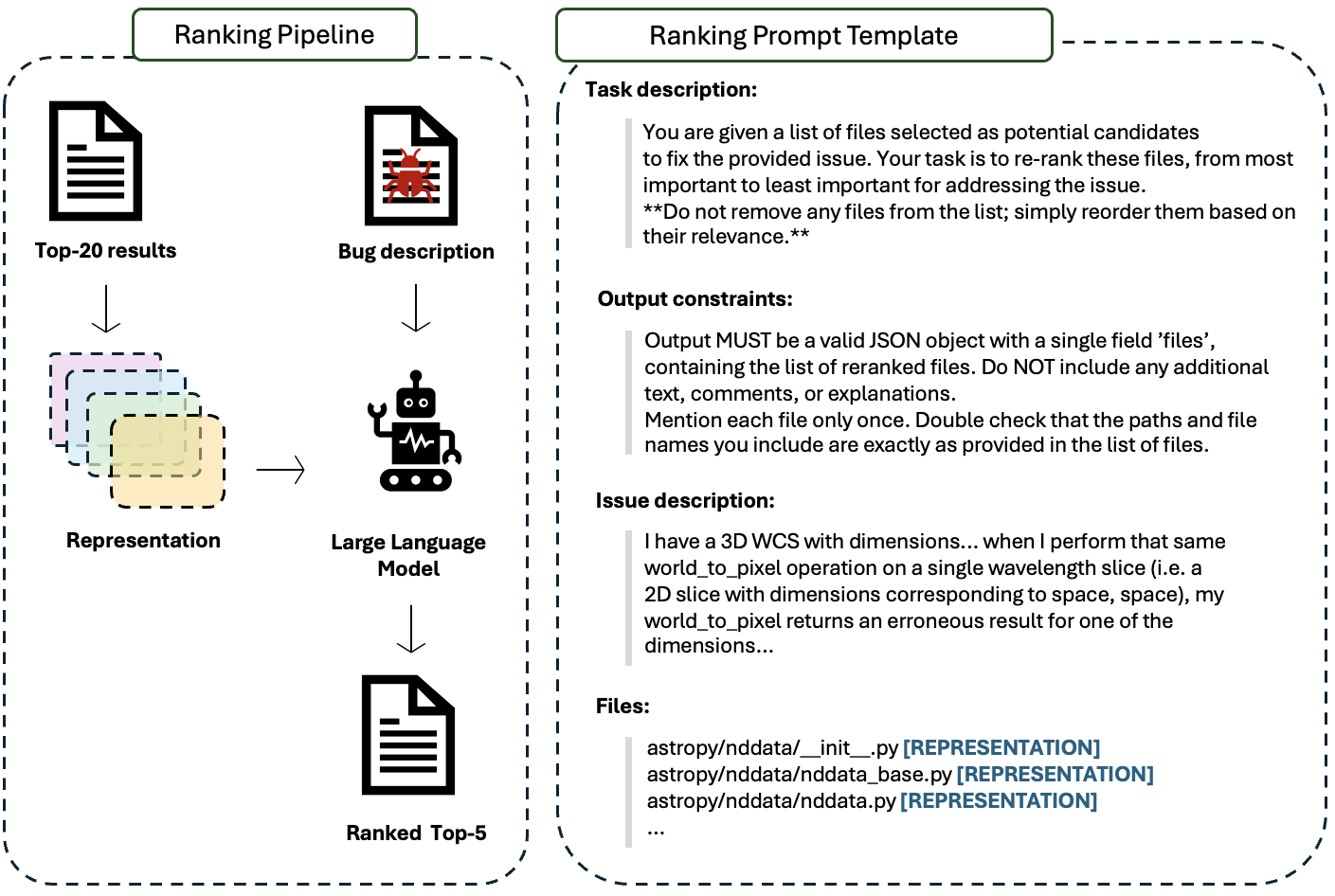}
    \caption{Sample Ranking Prompt Template. ``[REPRESENTATION]'' is a placeholder for any representation under study.}
    \label{fig:ranking_prompt_template}
\end{figure}
We show a sample prompt template and our ranking pipeline in Figure~\ref{fig:ranking_prompt_template}. As for the prompt template in Section~\ref{sec:rq2}, we divide our prompt in blocks: 1) task description, 2) output constraints, and 3) issue description and list of files. All representations include the file path, along with the chosen representation.
We prompt the LLM with our top-$k$ results and rank them using one code representation.

Ranking prompts inherit the same limitations as in RQ2: context size limits. 
We retain a context size limit of 16K for reasons of efficiency on our hardware. 
When the candidate set contains only one file, or when the representation budget permits only one candidate to be shown, the original ordering is retained. Invalid outputs from the LLM are treated as a failure, and the original ranked list is returned instead.

\subsection{Results}
We report the results of ranking per model in Table~\ref{tab:rq3_results_dense}, and summarize the main findings below.%
\vspace{0.5em}

\begin{table*}
    \centering    
    \footnotesize
    \setlength{\tabcolsep}{2pt}
    \caption{Impact of Ranking with Different Context Composition Strategies on dense embeddings. Percentages report the relative Hit$_5$ difference from the \noranking~baseline. Positive and negative differences are shown in \textcolor{green!60!black}{green} and \textcolor{red!60!black}{red}, respectively.}
    \resizebox{\textwidth}{!}{%
    \input{rq3_results_dense}
    }
    \label{tab:rq3_results_dense}
\end{table*}

\noindent
\textbf{Post-retrieval ranking improves over the \noranking~baseline for most representation-embedding pairs.} Table~\ref{tab:rq3_results_dense} shows that Qwen 3 peaks at 76.7\% Hit@5 with role-aware summaries on LCA, a 42\% increase against the baseline. It also reaches best ranking quality with a 79.7\% increase. Even if the baseline results were produced from the file path representation, it still yields improvements in a ranking setting: CodeXEmbed reaches maximum retrieval accuracy with file paths on LCA (+49.3\%) and on SWE (+39.1\%). On the LCA dataset, increases are consistent across the board. On SWE, Qwen 3 peaks at 0.788 Hit@5 (+25.9\%) with role-aware summaries. However, both models lose accuracy with raw source files. Our hypothesis is that raw sources are at a disadvantage given the limitation of one prompt to fit all representation context.
\vspace{0.5em}

\vspace{0.5em}

\noindent
\textbf{Ranking with bug report summaries (\bugreports) shows competitive performance with both datasets and models.} On LCA, MAP@5 is competitive with role-aware summaries, beating the baseline on both ranking quality and relevant retrieved files by up to 39.4\%. On SWE, bug report summaries beat the baseline, reaching a 31\% Hit@5 increase with CodeXEmbed and 16.9\% with Qwen 3.
\vspace{0.5em}

\noindent
\textbf{Raw sources (\rawsources) performance is limited in a single-prompt setting.} Raw source files are limited across the board by token budget constraints in a single-prompt ranking setting, while still improving against the baseline in some cases. We observe decreases of up to 5.4\%  against the baseline on the SWE dataset. 
\vspace{0.5em}

\subsubsection*{Files viewed per representation}

\begin{figure}
    \centering
    \includegraphics[width=1\linewidth]{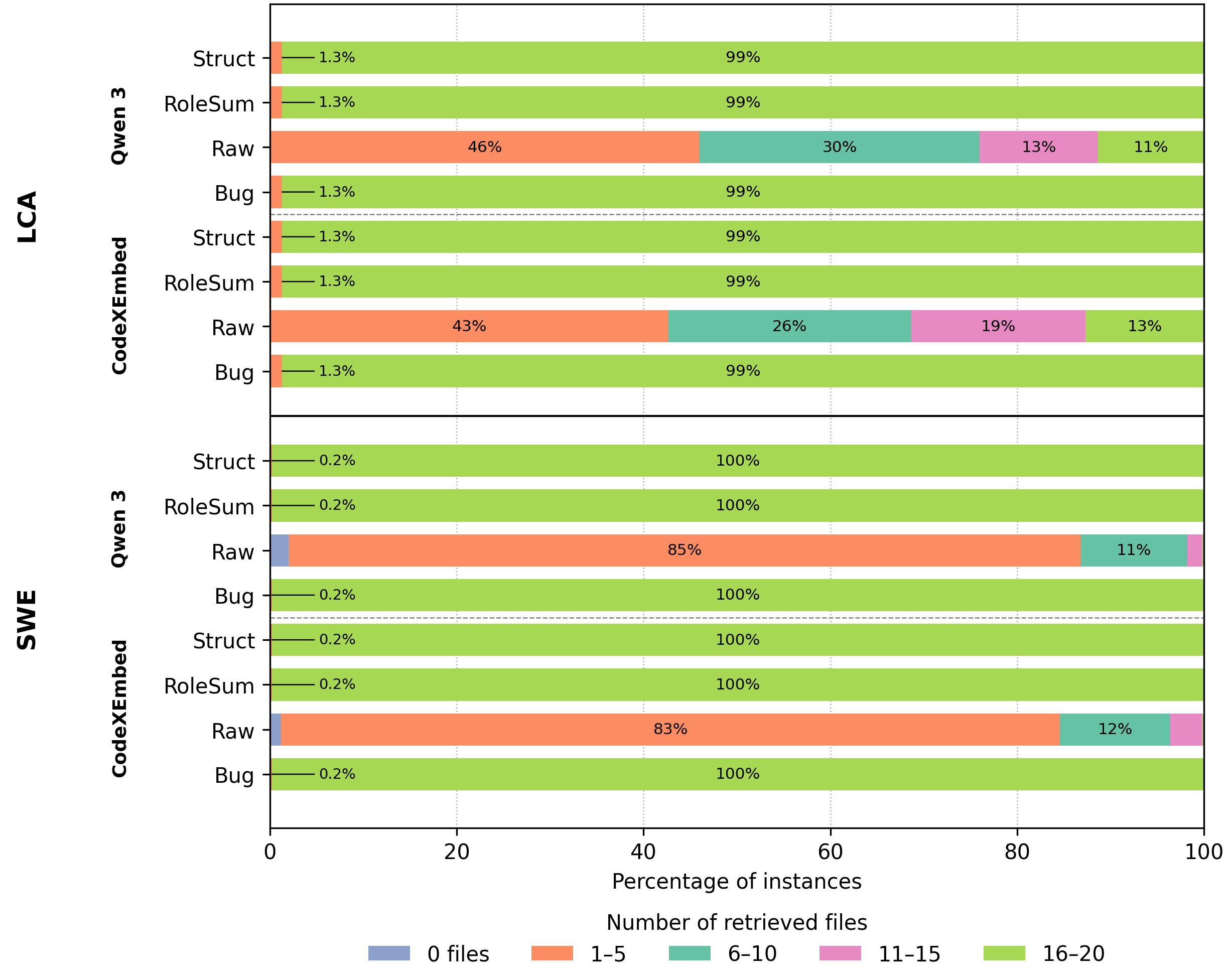}
    \caption{Distribution of the number of files viewed in a single prompt per representation and dataset. Path: file paths, Sum: role-aware summaries, Raw: raw source files, Bug: generated bug report summaries.}
    \label{fig:rq3_raw_distribution}
\end{figure}

Embedding models retrieve up to 20 candidate files, but ranking is performed using a single prompt and is constrained by the context window. To better understand representation limitations in single-prompt ranking and analyze the distribution of files showed to the ranker, we gather, for each task, how many files were included in the ranking prompt. We report the distribution of files viewed by the ranker for each representation in Figure~\ref{fig:rq3_raw_distribution}.
\vspace{0.5em}

\noindent
\textbf{The size of raw source (\rawsources) representation reduces ranking opportunity, particularly on SWE.} File paths, role-aware summaries, and generated bug reports provide a median of 20 viewed candidates within one prompt on both datasets. On the other hand, raw sources limit the number of viewed files: only 1-5 files are shown for 46\% and 43\% of instances on LCA for Qwen 3 and CodeXEmbed, respectively. 
On SWE, 85\% and 83\% of instances include only 1-5 files in the context for those models. On SWE, we observe instances where no files are shown to the ranker, indicating that the ranker encountered an error and returned no files.
\vspace{0.5em}

\vspace{0.5em}

\noindent
\textbf{Lightweight representations preserve the whole candidate set within a single prompt.}
File paths, role-aware summaries and generated bug reports expose a median of 20 candidates for both embedding retrievers on both datasets. The small proportion of single-file contexts, ranging from 0.2\% to 1.3\%, arises from dataset instances containing only one source file in the target programming language, rather than from retrieval or context-window limitations. This shows that lighter representations compress information in an effective, more scalable manner.

\vspace{1em}

\begin{tcolorbox}[rqbox]
\textcolor{paleblue!30!black}
{\textbf{Takeaway.} Ranking consistently improves results, even when using the same representation as the retriever. Lighter representations are a good option for single-prompt ranking settings, while raw source show limitations due to their large footprint. Generated bug report summaries beat the baseline by up to 39.4\% on Hit@5, showing their utility in a ranking stage.
}
\end{tcolorbox}

%% file: rq3_results_dense.tex
\begin{tabular}{@{}l
    rr
    rr@{\,}>{\tiny}r
    rr@{\,}>{\tiny}r
    rr@{\,}>{\tiny}r
    rr@{\,}>{\tiny}r@{}}
\toprule
\multirow{2}{*}{\textbf{Model}}
& \multicolumn{2}{c}{\textbf{\noranking}}
& \multicolumn{3}{c}{\textbf{\projectstructure}}
& \multicolumn{3}{c}{\textbf{\filesummarytwo}}
& \multicolumn{3}{c}{\textbf{\rawsources}}
& \multicolumn{3}{c}{\textbf{\bugreports}} \\
\cmidrule(lr){2-3}
\cmidrule(lr){4-6}
\cmidrule(lr){7-9}
\cmidrule(lr){10-12}
\cmidrule(lr){13-15}

& \multicolumn{1}{c}{\textbf{MAP$_5$}}
& \multicolumn{1}{c}{\textbf{Hit$_5$}}

& \multicolumn{1}{c}{\textbf{MAP$_5$}}
& \multicolumn{1}{c}{\textbf{Hit$_5$}}
& \multicolumn{1}{c}{\tiny\textbf{\%}}

& \multicolumn{1}{c}{\textbf{MAP$_5$}}
& \multicolumn{1}{c}{\textbf{Hit$_5$}}
& \multicolumn{1}{c}{\tiny\textbf{\%}}

& \multicolumn{1}{c}{\textbf{MAP$_5$}}
& \multicolumn{1}{c}{\textbf{Hit$_5$}}
& \multicolumn{1}{c}{\tiny\textbf{\%}}

& \multicolumn{1}{c}{\textbf{MAP$_5$}}
& \multicolumn{1}{c}{\textbf{Hit$_5$}}
& \multicolumn{1}{c}{\tiny\textbf{\%}} \\
\midrule

\multicolumn{15}{l}{\textbf{LCA Dataset}} \\
\midrule

Qwen~3
& 0.227 & 0.540
& 0.390 & 0.753
  & \textcolor{green!60!black}{+39.4\%}
& \textbf{0.408} & \textbf{0.767}
  & \textcolor{green!60!black}{+42.0\%}
& 0.375 & 0.580
  & \textcolor{green!60!black}{+7.4\%}
& 0.405 & 0.753
  & \textcolor{green!60!black}{+39.4\%} \\

CodeXEmbed
& 0.234 & 0.487
& 0.378 & {0.727}
  & \textcolor{green!60!black}{+49.3\%}
& 0.378 & 0.667
  & \textcolor{green!60!black}{+37.0\%}
& 0.370 & 0.580
  & \textcolor{green!60!black}{+19.1\%}
& {0.379} & 0.667
  & \textcolor{green!60!black}{+37.0\%} \\

\midrule
\midrule

\multicolumn{15}{l}{\textbf{SWE Dataset}} \\
\midrule

Qwen~3
& 0.418 & 0.626
& \textbf{0.640} & 0.780
  & \textcolor{green!60!black}{+24.6\%}
& 0.639 & \textbf{0.788}
  & \textcolor{green!60!black}{+25.9\%}
& 0.435 & 0.592
  & \textcolor{red!60!black}{$-$5.4\%}
& 0.599 & 0.732
  & \textcolor{green!60!black}{+16.9\%} \\

CodeXEmbed
& 0.437 & 0.516
& {0.586} & {0.718}
  & \textcolor{green!60!black}{+39.1\%}
& 0.581 & 0.716
  & \textcolor{green!60!black}{+38.8\%}
& 0.370 & 0.504
  & \textcolor{red!60!black}{$-$2.3\%}
& 0.556 & 0.676
  & \textcolor{green!60!black}{+31.0\%} \\

\bottomrule
\end{tabular}

%% file: discussion.tex
\section{Discussion} \label{sec:discussion}

\subsection{Case Study: Impact of representations on \textit{Agentless}.} \label{sec:agentless_case_study}
To evaluate whether our representation findings transfer to an existing bug localization workflow, we apply them to \textit{Agentless}~\citep{xia2024agentlessdemystifyingllmbasedsoftware}. 
\textit{Agentless} implements an end-to-end repair workflow without autonomous agents, using a staged hierarchical localization pipeline that begins with file-level localization, then narrows to edit locations and patch generation.
We do not intend to reproduce the full repair pipeline: instead, we focus on the file-localization stage and ask whether the representations identified in our study can improve the candidate sets in a well-established workflow. The \textit{Agentless} localization pipeline first asks an LLM to identify relevant files from the project structure (file paths). It then asks the LLM to identify irrelevant folders, and finally performs embedding-based retrieval with the text-embedding-3-small dense embedding model while excluding those folders. The resulting candidate lists are combined using a voting-based strategy. 

In our experiments, we follow this overall structure, but vary the representations used during localization: we integrate our most cost-effective generated representation, role-aware summaries (\filesummarytwo), and compare it with file paths and embedding retrieval over raw sources. We use the reciprocal rank fusion algorithm selected in our study for combining all results. 

Guided by the results of our study, we evaluate two adaptations of the \textit{Agentless} file localization stage:

\begin{enumerate} 
    \item \textit{Agentless-w-role-summaries}: combining file paths retrieval, role-aware-summary retrieval, and embedding retrieval over raw source files; 
    \item \textit{Agentless-best-w-ranking}: applying our LLM-based post-retrieval ranking settings with role-aware summaries to the best combined candidate set. 
\end{enumerate}

\textbf{Dataset.} We use SWE-bench Verified (SWE), a dataset used by both our study and \textit{Agentless}.

\textbf{Models.} We use the best performing model from Section~\ref{sec:rq2}, Qwen2.5-coder 32B. Baseline experiments from the original work are re-run with this model for comparability. We use text-embedding-3-small for embedding retrieval, following the \textit{Agentless} setup.

\textbf{Metrics.} As \textit{Agentless} reports the top-6 best candidates on combined results, we report Hit@6 both on individual representations and on combined results, for comparability. Each retrieval source contributes up to six candidates. Thus, when combining results, each model/representation pair adds one set of top-6 results, for a total of up to 18 candidates in the post-retrieval ranking stage.
\vspace{1em}

\subsubsection*{Results} 
We show the results of different combinations in Table~\ref{tab:agentless_results}. Role-aware summaries and file paths are used by Qwen2.5-coder 32B, while raw sources are used by text-embedding-3-small.
\vspace{0.5em}

\noindent
\textbf{Including role-aware summaries (\filesummarytwo) yields improvements.}
\textit{Agentless-w-role-summaries} yields a 2.2\% improvement over the baseline. Because both experiments use an identical combined retrieval logic and the same underlying model (Qwen2.5-coder 32B) as the baseline, this delta is attributed solely to the added representation. 
\vspace{0.5em}

\noindent 
\textbf{LLM-based post-retrieval ranking further improves the best candidate set.} Applying LLM-based post-retrieval ranking with role-aware summaries to the best combination increases Hit@6 to 0.940, a 4.7\% improvement over the baseline. This result supports one of the main findings of Section~\ref{sec:rq3} (RQ3): compact, semantically dense representations are useful when an LLM must compare several candidate files within a limited context budget.
\vspace{0.5em} 

Overall, this case study demonstrates in a practical way that combining representations improves file-level localization effectiveness. Role-aware summaries improve LLM-based retrieval, complement file paths and raw source retrieval, and provide an effective representation for post-retrieval ranking. 

\begin{table}
\centering
\setlength{\tabcolsep}{4pt}
\caption{Agentless file-level bug localization performance on SWE with our representation-aware pipeline stages. For each experiment, the percentage reports the relative difference in Hit@6 with the baseline, marked by $\dagger$.}
\label{tab:agentless_results}
\input{rq3_agentless_results}
\end{table}

\subsection{Representation and Retriever Costs}

\subsubsection*{Indexing Time} \label{sec:indexing_costs}
Different retriever types incur different indexing time costs. 
Dense embeddings have the added cost of generating embeddings and storing them in a vector database, while BM25 creates and stores an index beside the repository files.
To determine an upper-bound indexing time for different retrievers and representations on our datasets, we measure the indexing time of our largest project from Long Code Arena, based on the number of lines.  
The Trino Java project snapshot contains 988,722 lines, by far the largest among our datasets. 
All experiments are executed on our local machine. We gather timings by using the \texttt{Time} Python package and express them in minutes.

We report \textbf{indexing times} in minutes for BM25 and dense embeddings in Table~\ref{tab:trino_times}. 
We refer to indexing time as the time required to create an index on disk (BM25) or to process each representation instance, convert it to an embedding and store it in the vector database (dense embeddings). For summaries, we assume they are already generated and available on disk at this point. 
As expected, raw source files incur the highest retrieval cost in most cases, while file paths are the most lightweight.

\begin{table}
\centering
\caption{Indexing time per retriever and representation for the Trino repository (LCA task 8247).}
\input{trino_indexing_times}
\label{tab:trino_times}
\end{table}

\noindent
\textbf{BM25 is efficient regardless of the representation used.} Sparse retrieval is comparatively insensitive to representation size at this scale: even the raw-source representation can be indexed in only a few seconds. It makes BM25 an efficient alternative for leveraging raw sources files, a heavy but detailed representation that is less efficiently processed by dense models.
\vspace{0.5em}

\noindent
\textbf{Representation matters for dense embedding models.}
Qwen 3 embeddings and CodeXEmbed, with 0.6 and 0.4 billion parameters, respectively, show a noticeable increase in indexing times with raw source files (up to 27 minutes). It is not the case with file paths, where Qwen 3 and CodeXEmbed reach comparable performance with indexing times under 30 seconds.
\vspace{0.5em}

\noindent
\textbf{Summaries provide a middle ground between cost and information richness.} File paths incur less costs for dense representations, with 22–29 seconds indexing time, but it is also the least informative representation. Raw sources is the most detailed, but it is expensive to embed. Summary representations preserve more semantic information than file paths, while avoiding the indexing cost of raw source code, costing up to 7.5$\times$ less. More detailed summaries increase indexing time relative to shorter summaries, but remain substantially cheaper than raw source indexing in most cases.
\vspace{0.5em}

We note an important discrepancy between CodeXEmbed performance on detailed technical summaries when compared with other models. We investigated this behavior by reducing the input token limit and batch size, but the slowdown persisted. During these runs, the system showed signs of disk offloading, suggesting a model-specific memory-efficiency issue. We therefore interpret this result cautiously and avoid generalizing it to technical summaries. 

\subsubsection*{Cost of Generating Summaries}
\label{sec:summary_costs}
Our results indicate that role-aware summaries offer the best cost-effectiveness overall. At retrieval time, their representation footprint is up to $\approx$20$\times$ smaller than raw source files while preserving important signals for retrieval, obtaining the best performance in several cases. 

However, summary-based representations require a generation step, which we call \textit{preprocessing cost}. At Groq’s January 2026 rates for GPT-OSS 20B (\$0.075 / million tokens input, \$0.30 / million token output), processing all files in Long Code Arena (150 repos) for all summary types cost \$272.53 USD. The larger SWE dataset (500 repos) cost \$734.47, pushing the bill to a total of \$1007.10 USD. This corresponds to approximately \$1.82 per repository for LCA and \$1.47 per repository for SWE when all generated summary types are included. If only one summary representation is generated, the expected cost is proportionally lower. These costs should be interpreted as one-time preprocessing costs rather than per-query costs. Once summaries are generated, they can be cached and reused across localization queries, and only summaries for changed files need to be regenerated. This amortization is important in practice: real-world deployments typically index one or a small number of large repositories, and indexing can be performed offline. Additionally, our use of an API was motivated by the need to process hundreds of large repositories; for a single organization or project, a local model may be sufficient for summary generation.

Overall, summaries introduce an up-front cost, but they reduce downstream retrieval and ranking costs by producing compact, semantically dense representations. They are therefore most attractive in settings where repositories are queried repeatedly, where downstream LLM context is limited, or where candidate files must be compared within a fixed prompt budget. %

\subsection{Complementarity of Representations}
Our study shows that combining representations with reciprocal rank fusion often yields improved results. To understand the contributions of each representation, we conduct a study using our best model from RQs 1 and 2. First, we study the complementarity of representations with Qwen 3 embeddings, a traditional retriever. Next, we study  complementarity with our best LLM, Qwen-2.5 Coder 32B.

We gather statistics about how many bugs are localized in total, and how many bugs are uniquely localized by each representation. This gives us information about how combining different representations can potentially result in retrieval gains, as well as the maximum number of bugs covered by all representations combined. 

We focus on overall coverage, dominant contributors, and unique contributions across models and datasets. We illustrate contributions with UpSet plots: a horizontal histogram on the left shows the total bugs found by each representation, while the vertical histogram shows unique contributions per representation group. Connected dots indicate groups that localize common bugs, while single dots represent unique per-representation contributions.
\vspace{0.5em}

\begin{figure*}
\centering
    \centering
    \includegraphics[width=\linewidth]{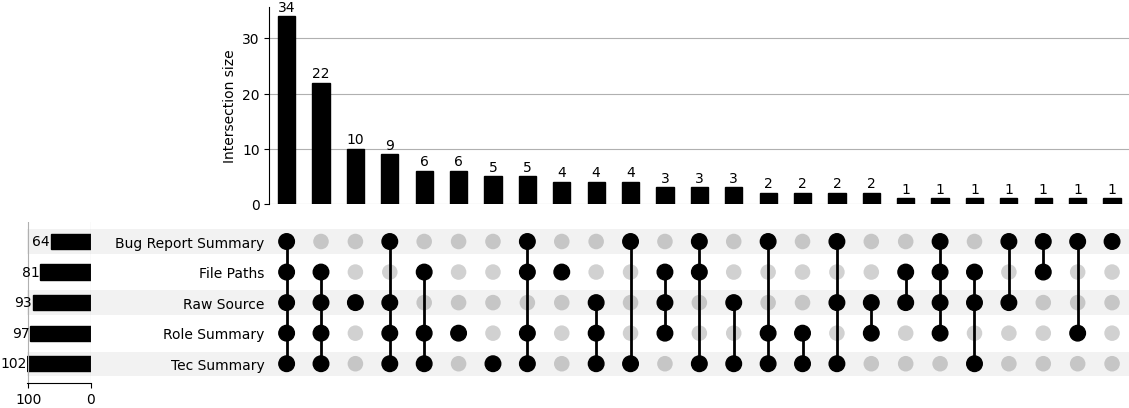}
    \caption{Overlap of bugs localized by different representations on LCA, showing the best performing model, Qwen 3. Union of representations = 133 bugs localized.}
\label{fig:upset_qwen3_model_lca}
\end{figure*}

\begin{figure*}
\centering
    \centering
    \includegraphics[width=\linewidth]{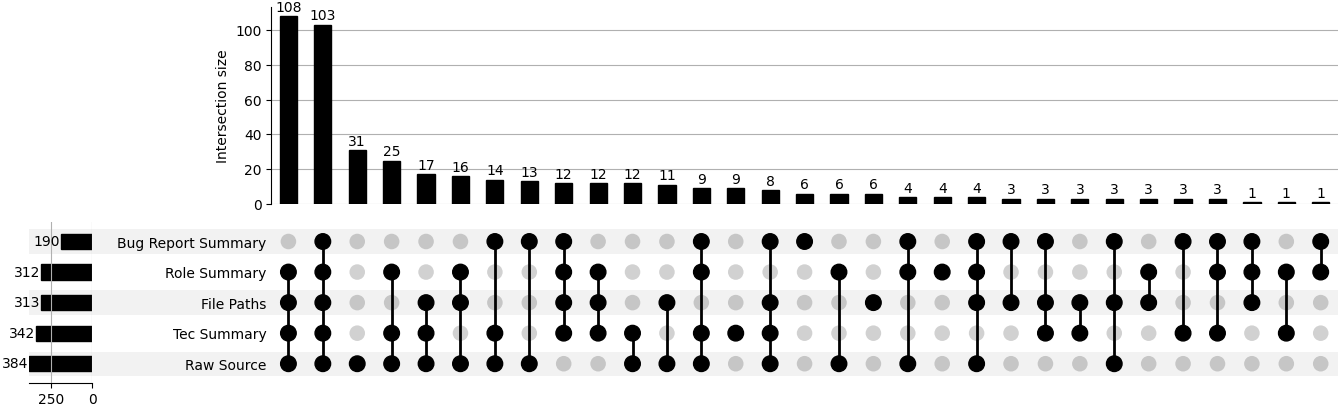}
    \caption{Overlap of bugs localized by different representations on SWE, showing the best performing model, Qwen 3. Union of representations = 454 bugs localized.}
\label{fig:upset_qwen_model_swe}
\end{figure*}

\subsubsection*{Complementarity with Qwen 3}

We show the complementarity of representations and their contributions with Qwen 3 embeddings for each dataset. Figure~\ref{fig:upset_qwen3_model_lca} shows the intersections between representations on LCA,
while Figure~\ref{fig:upset_qwen_model_swe} shows the results for SWE.
\vspace{0.5em}

\noindent
\textbf{Raw sources (\rawsources) make the most unique contributions.} 
On LCA, Figures~\ref{fig:upset_qwen3_model_lca} shows that raw sources makes the most unique contributions for Qwen 3 (10). On SWE, Figure~\ref{fig:upset_qwen_model_swe} shows that raw sources find 31 unique bugs. 
\vspace{0.5em}

\noindent
\textbf{File paths (\projectstructure) overlap heavily with richer representations}. Despite encoding only repository-relative paths and file names, this representation is a strong contributor on total bugs localized and finds up to 4 unique bugs on LCA and 6 on SWE. 
\vspace{0.5em}

\noindent
\textbf{Bug report summaries (\bugreports) localize few unique bugs.} They contribute only one unique localization with Qwen~3 on LCA. On SWE, all representations except bug report summaries are major contributors, overlapping on 108 bug instances with Qwen~3.
\vspace{0.5em}

\subsubsection*{Complementarity with Large Language Models}

\begin{figure}
    \centering
    \includegraphics[width=0.75\linewidth]{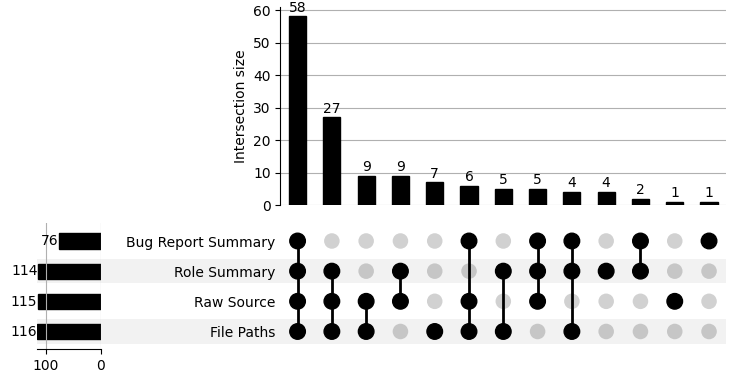}
    \caption{Qwen 32B - Representation contributions to localized bugs on LCA. Union of representations = 138 bugs localized.}
    \label{fig:upset_qwen32}
\end{figure}

\begin{figure}
    \centering
    \includegraphics[width=0.75\linewidth]{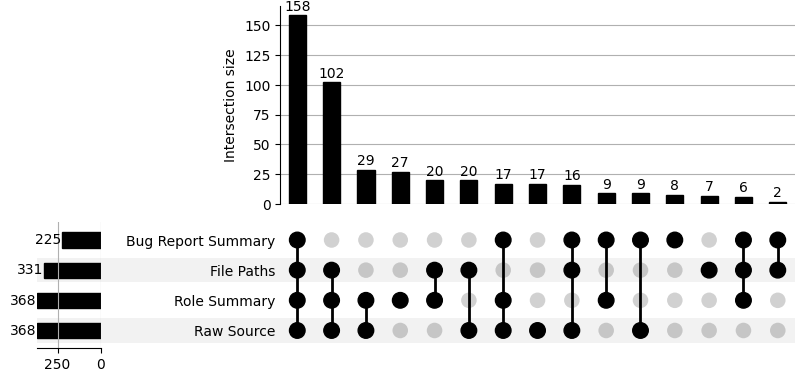}
    \caption{Qwen 32B - Representation contributions to localized bugs on SWE. Union of representations = 447 bugs localized.}
    \label{fig:upset_qwen32_swe}
\end{figure}

We show the complementarity of representations and their contributions with our best LLMs per dataset. Figure~\ref{fig:upset_qwen32} shows the intersections between representations on LCA, while figure~\ref{fig:upset_qwen32_swe} shows the results for SWE.
\vspace{0.5em}

\noindent
\textbf{A large overlap is observed between representations.} Qwen 32B results shows that all representations contribute 58 common bugs localized on LCA (Figure~\ref{fig:upset_qwen32}), while individual unique contributions remain low with the highest contributor, file paths, bringing in only 7 unique findings. On SWE the trend is similar: Figure~\ref{fig:upset_qwen32_swe} show 158 contributions by all representations, and only 27 unique contributions by role-aware summaries.
\vspace{0.5em}

\noindent
\textbf{Role-aware summaries (\filesummarytwo) and raw sources (\rawsources) localize the most bugs on SWE.} Raw sources and role-aware summaries both localize 368 out of 500 bugs on SWE, followed by file paths (331). Role-aware summaries identifies the most unique bugs with this dataset (27). On LCA, file paths, raw sources and role-aware summaries are competitive with 116, 115 and 114 bugs localized, respectively.
\vspace{1em}

\subsubsection*{Example of Representation Complementarity}
In this section, we aim to show how the role-aware summary representation can uniquely localize issues that could not be localized by any other representation. To ease interpretation, we select a sample retrieved with BM25, where lexical overlap between the issue and the indexed representation is easier to inspect. We have shown in Section~\ref{sec:rq1} that BM25 generally performs better with raw source files. Our example shows that another representation, even if not the best performing, can be complementary.

We identify the bug \texttt{astropy-13579} from the SWE-bench Verified dataset, where the single ground truth file is localized by the generated role-aware summary (\filesummarytwo) at rank~\#3 of the top-5 results. Figure~\ref{fig:compl-snippet} shows an excerpt of the issue description and the role-aware summary that localized the buggy file.

\input{complementarity_snippet}

The issue excerpt and reproduction code contain many relevant code identifiers that appear in both the raw source file and the role-aware summary. The role-aware summary is effective because of the high density of relevant identifiers in a compact context: WCS, slice, pixel/world dimensionality, class name, etc. The raw source contains the same identifiers, but they are diluted across a long implementation and mixed with lower level control flow and array manipulation logic. Thus, the summary improves retrievability by placing the identifiers in a compact description of the file's role and responsibility.

\subsection{Implications}

\subsubsection*{Insights for Practitioners}
Our findings show that how context is composed matters as much as the retriever itself. 
Different representations expose different retrieval signals and incur different costs, so the best configuration depends on the practitioner’s accuracy, latency, and budget constraints. For lightweight first-stage retrieval, raw source code remains a strong option when paired with BM25. In our experiments, BM25 over raw sources provides a favourable accuracy-cost trade-off: it avoids the generation and indexing costs of summaries while preserving the lexical identifiers, API calls, and implementation details that often appear in bug reports. Combining a lexical retriever over raw sources with a dense one over complementary representations, such as file paths or summaries, can further improve coverage when additional storage and indexing cost are acceptable. Previous works have acknowledged the efficiency of hybrid retrieval pipelines~\citep{gao_complement,lu2022reacc,contextual_retrieval}.

However, raw sources are less suitable as direct input to LLM-based retrieval or ranking across many candidate files. Their large footprint increases latency and context-window pressure, and it can reduce the number of files that fit into the prompt. In these settings, compact representations such as role-aware summaries provide a more practical way to expose repository-level context to LLMs. They concentrate the file’s responsibilities, APIs, and likely failure modes into fewer tokens, making them useful when the model must compare multiple candidates within a limited context window.

Practitioners should also account for model-specific behavior. Smaller or less capable models may benefit from more explicit lexical and structural cues, while larger models can exploit summaries more effectively. As a result, representation choice should be tuned jointly with the retrieval model and the deployment constraints. A practical strategy is to use inexpensive lexical retrieval over raw sources to generate candidates, then apply summaries or other compact representations for ranking or downstream LLM reasoning when higher precision is needed.

Our results also show that LLMs can be effective retrievers, especially when using the larger 32B model. However, this effectiveness comes at a higher cost, and unlike BM25 or dense retrieval, which can reliably return a fixed top-$k$ list, LLM-based retrieval sometimes returns only a small number of candidates. For instance, a single file is returned in 48/150 occurrences on LCA, and 309/500 for SWE. The high rate of single-file predictions on SWE (61.8\%) suggests that the LLM is conservative when retrieving relevant files. This limits downstream ranking or inspection stages that require a sufficiently large candidate set. Practitioners should therefore treat LLM-based retrieval as a higher-cost, precision-oriented component rather than as a direct replacement for inexpensive first-stage retrieval.

\vspace{0.5em}

\textbf{Usability in Development Workflows.}
The experiments in this study can all be run on accessible consumer hardware, given enough random access memory is available. Larger models require buying specialized, expensive hardware that is not easily accessible to any research laboratory or company. Another option is to access models through an API, but this comes with limitations: 1) Models can be deprecated at any time; 2) model versions can change; 3) recurring costs, which will change in the future, need to be factored into development costs; 4) larger models can not be fine-tuned to specific tasks due to their large scale~\citep{Huyen_2024_evalai}.  Using smaller models solves most of those issues, but they have their own drawbacks which we explain below.
\vspace{0.5em}

\textbf{Challenges of using smaller models.} Working with smaller LLMs requires more prompt engineering and different strategies to increase accuracy~\citep{swe-bench}. 
In our experiments, several adjustments were required to get better-structured outputs. Generating structured outputs is a challenge for smaller models, in particular our 7B-parameter model. Post-processing has to be done more carefully, as we can rarely read in a JSON object in one go with no errors. Smaller models make mistakes when copying file paths from the input prompt, which causes our pipeline to reject candidates. For instance, they sometimes prepend a slash to the relative path, add an extra space, etc. These failures are consistent with recent evidence that instruction following ability impacts task performance. ~\citet{murthy2025kcifknowledgeconditionedinstructionfollowing} introduce KCIF, a benchmark that augments knowledge tasks with simple answer-modifying and distractor instructions, and show that these additional constraints can substantially reduce performance especially in small and medium models. In our setting, the model must not only identify relevant files, but also follow formatting, copying, and output-structure constraints. 

The above points add complexity to the pipeline. Smaller models may benefit from task-specific fine-tuning, as suggested by~\citet{swe-bench}.

\subsubsection*{Insights for Researchers}
Our results suggest that code representations should be studied as part of the retrieval architecture, rather than as interchangeable inputs to a single retriever. Prior work has explored coarse-to-fine workflows that progressively narrow the search space before exposing detailed code to an LLM~\citep{flexfl,xia2024agentlessdemystifyingllmbasedsoftware}. Our findings complement this direction by showing that the representation used at each stage changes both the information available to the retriever and the amount of context available to downstream models. In particular, combining representation results can improve top-5 coverage by combining complementary signals, while compact representations can increase the number of candidate files that fit within an LLM ranking prompt. This matters for agentic workflows because downstream reasoning, edit localization, and dependency analysis can only operate over the candidates that are retrieved in the top-$k$. 

Our results also point to a broader opportunity for hybrid retrieval architectures. Combining multiple representations can improve localization, but further gains may require combining multiple retrievers as well. Prior work in information retrieval has shown that lexical and neural retrieval signals are complementary, with hybrid systems combining BM25-style exact matching and semantic retrieval often outperforming either signal alone~\citep{rrf,gao_complement}. Similar observations appear in bug localization, where combining textual similarity, structured source-code information, version history, similar reports, and learned representations improve localization effectiveness~\citep{bluir,wang_combine,bug_loc_deep_learning}. Future research should evaluate representation fusion and retriever fusion jointly, while reporting not only localization accuracy but also representation footprint, context exposure, and the number of candidates available to downstream models.

%% file: rq3_agentless_results.tex
\newcommand{\baseline}{\textsuperscript{\(\dagger\)}}
\newcommand{\updelta}[1]{\textcolor{green!60!black}{(+#1\%)}}
\newcommand{\downdelta}[1]{\textcolor{red!70!black}{(#1\%)}}

\begin{threeparttable}
\begin{tabular}{l r@{\,}>{\tiny}r l}
\toprule
\textbf{Configuration} & \multicolumn{2}{c}{\textbf{Hit@6}} & \textbf{Representations}\\
\midrule
\textit{Agentless}\baseline\ 
    & 0.898 &  & \projectstructure\rawsources \\
\textit{Agentless-w-role-summaries}
    & 0.918 & \updelta{2.2} & \projectstructure\filesummarytwo\rawsources\\
\midrule
\textit{Agentless-best-w-ranking}
    & \textbf{0.940} & \updelta{4.7} &\projectstructure\filesummarytwo\rawsources\\
\bottomrule
\end{tabular}

\begin{tablenotes}
\footnotesize
\item \(\dagger\): Baseline. 
\item File paths and role-aware summaries used with Qwen2.5-coder 32B, raw sources used with text-embedding-3-small.
\end{tablenotes}
\end{threeparttable}

%% file: trino_indexing_times.tex
\begin{tabular}{@{}l rrrrr@{}}
\toprule
\multirow{2}{*}{\textbf{Retriever}}
& \multicolumn{5}{c}{\textbf{Indexing Time (minutes)}} \\
\cmidrule(lr){2-6}\\

& \projectstructure
& \rawsources
& \filesummarytwo
& \filesummarythree
& \bugreports \\

\midrule
BM25 & 0.06 & 0.11 & 0.07 & 0.09 & 0.06 \\
GTE-Large & 0.38 & 17.40  & 2.41 & 4.11 & 1.94\\
Qwen 3 & 0.45 & 26.97 & 4.00 & 4.02 & 2.91\\
CodeXEmbed & 0.49 & 24.42 & 3.27 & *48.51 & 2.65\\
BGE-Large & 0.37 & 17.28 & 2.36 & 4.11 & 1.93\\

\bottomrule

\multicolumn{6}{@{}l@{}}{
\footnotesize *CodeXEmbed performance on technical summaries is an outlier,
}\\
\multicolumn{6}{@{}l@{}}{
\footnotesize due to memory efficiency issues of the model on our system.
}\\
\end{tabular}

%% file: complementarity_snippet.tex
\begin{figure}[t]
\caption{Example of a bug localized only by a role-aware summary using BM25. The issue excerpt is shortened for readability. Terms in the summary that overlap with the issue excerpt are highlighted.}
\label{fig:compl-snippet}

\footnotesize
\setlength{\tabcolsep}{4pt}
\begin{tabularx}{\columnwidth}{@{}p{0.18\columnwidth}X@{}}
\toprule
\textbf{Issue excerpt} &
[...] I find that when I perform a \texttt{\hl{world\_to\_pixel}} on the full (unsliced) \hl{WCS}, I get back the expected result. However, when I perform that same \texttt{\hl{world\_to\_pixel}} operation on a single wavelength \hl{slice} (i.e. a 2D \hl{slice} with \hl{dimensions} corresponding to space, space), my \texttt{\hl{world\_to\_pixel}} returns an erroneous result for one of the \hl{dimensions}. [...] it seems to be an issue with \texttt{\hl{SlicedLowLevelWCS}} [...]
\\
 & 
\begin{tabular}[t]{@{}l@{}} \texttt{ll\_sliced\_wcs = \hl{astropy.wcs}.wcsapi.\hl{SlicedLowLevelWCS}(fits\_wcs, 0)} \\ \texttt{hl\_sliced\_wcs = \hl{astropy.wcs}.wcsapi.HighLevelWCSWrapper(ll\_sliced\_wcs)} \\ \texttt{hl\_sliced\_wcs.\hl{world\_to\_pixel}(pt)} \\ \texttt{(\hl{array}(1.81818182e+11), \hl{array}(12.))} \end{tabular}

\\ 

\end{tabularx}

\footnotesize
\setlength{\tabcolsep}{3pt}
\begin{tabularx}{\columnwidth}{p{0.18\columnwidth}X@{}}
\toprule
\textbf{Retrieved file} & \textbf{Role-aware summary} \\
\midrule
\texttt{sliced\_wcs.py} &

\texttt{\hl{SlicedLowLevelWCS}} is a wrapper for astropy's \texttt{BaseLowLevelWCS} that applies an array \hl{slice} to a \hl{WCS} without modifying the underlying object. It sanitises slice inputs, combines nested slices, and tracks which \hl{pixel} and \hl{world} \hl{dimensions} are kept or dropped. The class exposes properties for \hl{pixel}/\hl{world} dimensionality, axis names, units, and correlation matrices, and implements \texttt{pixel\_to\_world\_values} and \texttt{\hl{world\_to\_pixel\_values}} that adjust for the \hl{slice} offsets. It also provides metadata about dropped world \hl{dimensions} and supports broadcasting of \hl{array} shapes. This module is used in the \texttt{\hl{astropy.wcs}} API to enable slicing of \hl{WCS} objects while preserving correct coordinate transformations.
\\

\bottomrule
\end{tabularx}

\end{figure}

%% file: threats.tex
\section{Threats to Validity}\label{sec:threats}

\noindent
\textbf{Output variability.} Large language models remain slightly non-deterministic, even when the temperature is set to 0. Ideally we would execute every experiment multiple times, but the full repository-scale runs were limited by time and compute budget.  
To keep the residual variance small we set the temperature to 0, and log every intermediate ranking and model response; these artifacts are released with the paper.  

\noindent
\textbf{File summaries generation.} File summaries' quality highly affects retrieval quality. The model used to generate the file summaries can significantly impact the results. For this study, due to the computational and monetary implications, we generated summaries using GPT-OSS 20B, available via the Grok API.

\noindent
\textbf{Metrics.} This threat relates to whether the metrics accurately measure the performance of retrieval for bug localization. We use widely accepted metrics (MAP and Hit@k), which are commonly used in previous studies. However, different development practices could influence the effectiveness of our evaluation.

\noindent
\textbf{Duplicates in embeddings retrieval.} For a single file, there may be multiple embeddings, depending on the model sequence length. Thus, a similarity search might return several chunks of the same file. As the results are de-duplicated, we may obtain fewer than k results than what was originally requested, which can impact the extracted MAP and Hit@k metrics. We mitigate this threat by retrieving all matches, then de-duplicating the results and finally, retrieving the top-5 results.

\noindent
\textbf{Dataset contamination.}
The LLMs we evaluate are pretrained on web-scale code corpora that may include the projects used in LCA and SWE-bench. Observed gains could stem from memorization rather than effective retrieval, and should be interpreted as an upper bound of real-world performance. Given that our experiment is meant to compare context composition strategies and their relative impact on the model's performance, we believe the results will hold.

\noindent
\textbf{Ollama-related crashes.} After a required update of Ollama, ten instances of SWE-bench crash the Ollama server when running the Files Summaries experiment for our replication. We record these as empty results, which likely underestimates true recall.

\noindent
\textbf{Generalizability.} Our experiments were run with three different models of sizes 7-32B on two large datasets spanning three programming languages: Java, Python and Kotlin. Our results already show a large variation of the context composition strategies across models, but we expect that the results may not hold for commercial code, other programming languages, or bug trackers with different writing styles.

\noindent
\textbf{Small LLM-retrieved candidate set.} In our experiments, the behavior of LLM-based retrievers raises an evaluation issue. The 32B model achieved strong localization performance, but it often returned very few candidate files. This makes standard top-$k$ comparisons incomplete: a retriever that returns one highly ranked file may perform well on Hit@$k$, but provides little opportunity for downstream ranking. For this reason, we did not use LLM-retrieved candidates as input to Section~\ref{sec:rq3}.

%% file: conclusion.tex
\section{Conclusion}\label{sec:conclusion}
We study how textual code representations affect file-level retrieval for agentic bug localization. Our findings indicate that no single representation dominates across models, datasets, and workflow stages. Raw source code can provide strong retrieval signals, but its large footprint makes it costly and difficult to use in ranking contexts. File paths are lightweight and often useful as a baseline, but lack semantic detail. Generated bug report summaries can capture failure-oriented descriptions, but their performance is inconsistent. Role-aware summaries provide the most robust cost-effectiveness tradeoff: they compress source files into a compact representation that preserves file responsibilities and improves alignment with natural-language bug reports. Our case study with \textit{Agentless} strengthens the practical applicability of our techniques, yielding improvements at the file localization stage when including role-aware summaries and post-retrieval ranking. These results suggest that repository-level localization workflows should treat representation as a design choice. Our study shows that improving agentic bug localization requires not only better models or retrieval algorithms, but also better control over the textual views of code that are made available to them.